\documentclass[a4paper,fleqn,usenatbib]{mnras}

\usepackage{newtxtext,newtxmath}
\usepackage[T1]{fontenc}
\usepackage{ae,aecompl}

\usepackage{graphicx}	% Including figure files
\usepackage{amsmath}	% Advanced maths commands
\usepackage{amssymb}	% Extra maths symbols
\usepackage{multirow}
\usepackage{wasysym}
\title[On the composition of moons and asteroids]{The composition of Solar system  asteroids and Earth/Mars moons, and the Earth-Moon composition similarity}

\author[Mastrobuono-Battisti \& Perets]{
Alessandra Mastrobuono-Battisti$^{1}$\thanks{E-mail: mastrobuono@mpia.de}
and Hagai B. Perets$^{2}$
\\
$^{1}$Max-Planck Instiut f\"ur Astronomie, K\"onigstuhl 17, D-69117 Heidelberg, Germany\\
$^{2}$Department of Physics, Technion, Israel Institute of Technology,
Haifa, 32000, Israel\\
}

\date{Accepted XXX. Received YYY; in original form ZZZ}

\pubyear{2016}

\begin{document}
\label{firstpage}
\pagerange{\pageref{firstpage}--\pageref{lastpage}}
\maketitle

% Abstract of the paper

\begin{abstract}
In the typical giant-impact scenario for the formation of the Moon, most of the Moon's material originates from the impactor. Any Earth-impactor composition difference should, therefore, correspond to a comparable Earth-Moon composition difference. Analysis of Moon rocks shows a close Earth-Moon composition similarity, posing a challenge for the giant-impact scenario, given that impactors were thought to significantly differ in composition from the planets they impact. However, our recent analysis of 40 planet formation simulations has shown that the composition of impactors could be very similar to that of the planets they impact. Moreover, the difference in Earth-Moon--like systems is consistent with observations of $\Delta^{17}O<15$ppm  for a significant fraction of the cases, thereby potentially resolving the composition similarity challenge. Here we use a larger set of 140 simulations and improved statistical analysis to further explore this issue.  We find that in 4.9\%-18.2\% (1.9\%-6.7\%) of the cases the resulting composition of the Moon is consistent with observations of $\Delta^{17}O<15$ppm ($\Delta^{17}O<6$ ppm).  These findings reaffirm our original results at higher statistical level and suggest that the Earth-Moon composition similarity could be resolved to  arise from the primordial Earth-impactor composition similarity. Note that although we find the likelihood for the suggested competing model of very high mass-ratio impacts (producing significant Earth-impactor composition mixing) to be comparable ($<6.7\%$), this scenario also requires additional fine-tuned requirements of a very fast spinning Earth. Using the same simulations we also explore the composition of giant-impact formed Mars' moons as well as Vesta-like asteroids. We predict that the Mars-moon composition difference should be large, but smaller than expected if the moons are captured asteroids. Finally, we find that the left-over planetesimals (`asteroids') in our simulations are frequently scattered far away from their initial positions, thus potentially explaining the mismatch between the current position and composition of the Vesta asteroid.
\end{abstract}

\begin{keywords}
Moon -- Earth -- planets and satellites: composition -- planets and satellites: formation -- planets and satellites: terrestrial planets -- minor planets, asteroids: individual: Vesta
\end{keywords}

\section{Introduction}

The Moon is thought to be the result of a relatively low-velocity, oblique impact \citep{Ca01} between the proto-Earth and a Mars sized
planetary embryo, Theia,  that happened in the latest stages of the 
Earth assembly  \citep{ACL99,JM14}.\\
Precise isotopic measurement of lunar rocks, brought back by the Apollo missions, showed that the Moon has an isotopic composition
almost indistinguishable from that of the Earth for various elements \citep[O, Ti, Cr, W, K;][]{Ri86,Lu98,Wi01,TK07,ZD12,HP14,2016Sci...351..493Y,2005AREPS..33..531J,TK07}. 
In particular, the oxygen isotope composition is characterised by a mutual difference $\Delta^{17}O=12\pm3$ppm
\citep{HP14}
or possibly smaller ($-1\pm5$ppm), as recently found by  \cite{2016Sci...351..493Y} analysing different rock samples. 
However, since objects in the Solar System have significantly different compositions (e.g. Mars, Vesta),  it is difficult to explain the Earth-Moon similarity in the giant-impact scenario in which most of the Moon material originates from the impactor.
Solutions to this conundrum through strong mixing during the encounter \citep{Cu12, Ca12} require many ad-hoc 
assumptions for the impact, in particular extremely high spin for the proto-Earth, and fine-tuned conditions \citep{JM14}.
High angular momentum impacts could explain the potassium chemical and isotopic composition of the Moon as shown by \cite{Wa16}.
%%%%%%%%
{
The origins of the Solar system structure, and in particular the low 
mass of Mars are still debated. Making use of one of the suggested 
scenarios which may successfully explain Mars origin (Grand Tack; \citealt{Wal11}; but see difficulties with the model, e.g. \citealt{Dan+12, 
Bro+17}), \cite{2016Sci...351..493Y}  found that the proto-Earth and Theia are 
likely to differ at the 0.1\permil~level, and they therefore also suggested 
that  a strong mixing following a high energy, high angular momentum 
impact in required to explain
the Earth-Moon composition similarity. However, as recently
shown by \cite{Ruf17}, such high spin is highly unlikely
because angular momentum is efficiently removed through material
ejected during the impact. Other models for low-mass mars, not involving 
migration \citep[e.g.][]{Bro+17} are yet to be tested using Moon formation 
models. It is also possible that a a paradigm shift in regard to the 
Moon origin is required.
}\\
{ Using elements with distinct affinities for metals to trace the composition of material accreting to form the Earth
at different times, \cite{Dau17} found that the Earth and Theia probably had similar chemical composition, thereby relaxing the constraints on the giant impact scenario.}\\
%%%%%%%%
The probability of an impact between two similar bodies, in the standard giant-impact scenario, has been explored 
using simulations of rocky planets formation. 
Assuming that the composition is a function of the initial distance of  planetesimals and planetary embryos from the Sun and 
analysing  the impact history of each planet, it is possible to obtain the mass-weighted composition of planets and the 
last impactors that could have formed the Moon. One can then compare the Earth and the Moon compositions in the simulations \citep[see][]{2015Icar..252..161K, 2015Natur.520..212M,  2015Icar..258...14K}. \\
{ Here we extend previous findings using a larger set of simulations
and applying an improved statistical analysis similar to the one described
by \cite{2015Icar..252..161K}
and \cite{2015Icar..258...14K}, but extend it as to now 
also consider the statistical errors, in order to better asses the composition
similarities of the Earth and the Moon. Such analysis can later be extended to Solar system and moon formation models beyond those considered here.
}

Moreover, we study the composition of Mars and its moons, and explore the composition of Solar system asteroids such as Vesta.
Mars moons Phobos and Deimos have indeed been reconsidered as
the result of a giant impact \citep{Cra11,Cit15} and Vesta is another interesting and peculiar object in our Solar System, 
showing a large discrepancy between its distance to the Sun and its composition relative to the Earth and Mars.  \\
The paper is organised as follows: in Section \ref{sec:methods} we describe the simulations and the analysis we performed.
Our results are illustrated Section \ref{sec:results} and in Section \ref{sec:concl}
we summarise the main outcomes and draw our conclusions.

\section{Methods} \label{sec:methods}
\subsection{$N$-body simulations}
Detailed $N$-body simulations of rocky planets formation have been recently used 
to study the composition of the Moon  \citep{PS07,2015Natur.520..212M, 2015Icar..252..161K, 2015Icar..258...14K, 2016Sci...351..493Y}.
Analysing only one simulation with 150 particles and considering all the impactors on every planet \cite{PS07} 
concluded that  their composition was too different to explain the Earth-Moon similarity.
In \cite{2015Natur.520..212M}, hereafter Paper I, we used a suite of 40 
$N$-body simulations  with different initial conditions \cite[see][]{Ra09},
to compare the composition of the planets formed with that of the relative last
massive impactor. Our analysis led to a probability of compatibility between the Earth and Theia 
between 20\% and 40\%, depending on the degree of mixing
allowed (from 0\% to 40\% of proto-Earth
material contributing to the circumterrestrial disk).
This hints to a general tendency for impactors and targets to have more similar compositions than different planets.
\cite{2015Icar..252..161K} used another set of
150 similar simulations and, with an analogous analysis, they obtained less than 5\%
chances to have an impact between similar bodies. Including the possibility
of mixing and a statistical analysis based on bootstrapping, \cite{2015Icar..258...14K} found that
the fraction of Theia analogs consistent with the canonical giant impact hypothesis is in the 5\%-8\% range.
The differences between the results obtained in these papers are mostly due to the choice of the analogs.
In Paper I the order of the planets has been used to identify the
Earth and Mars analogs (the third and forth planet in the system, respectively); moreover all
the planets and last impactors  where taken into account. 
\cite{2015Icar..252..161K, 2015Icar..258...14K} adopted the identification criteria
described in Section \ref{susect:analogs} and considered only the impactors on Earth's analogs.\\

We used the set of 40 simulations (set I) used in Paper I combined with
additional 100 simulations (set II) with similar initial conditions
analysed in \cite{2015Icar..252..161K, 2015Icar..258...14K} (this  latter set of simulations is publicly available).
The simulations, performed using Mercury code \citep{Ch99}, reproduce
the latest stages of planetary formation following the birth of Jupiter
and Saturn, after all the gas has been dissipated from the system
or accreted by the giants \citep[see][for a recent review]{Ra13}.
Set I and II include respectively eight and two ensembles with different
initial conditions. 

Simulations in set I begin from a disk of 85-90 planetary embryos and
1000-2000 planetesimals distributed between $0.5$ and $4.5$~au \citep[see][for more details]{Ra09}. In brief, in 
12 simulations 
%, named cjs and cjsecc, 
Jupiter and Saturn are mutually inclined with an angle of
$0.5^{\circ}$ and orbit the Sun with 
semi-major axes of $a_J=5.45$~au for Jupiter and $a_S=8.18$~au for Saturn. In 8 of those simulations the orbits 
are circular {(cjs)}, while in the other 4 they
are slightly  eccentric with $e_{J}=0.02$ for Jupiter and $e_{S}=0.03$ for Saturn {(cjsecc)}. In 12 additional simulations %, eejs, 
Jupiter and Saturn are located in their current positions (5.25 and 9.54 au),
with mutual inclination of $1.5^{\circ}$ and eccentricities
$e_{J}=e_{S}=0.1$ or $e_{J}=0.07$ and $e_{S}=0.08$ { (eejs)}.
%In ejs 
In 8 runs \cite{Ra09} adopted orbital parameters similar to the observed ones ($a_{J}=5.25$au and $e_{J}=0.05$, $a_{S}=9.54$ and
$e_{S}=0.06$) with mutual inclination of $1.5^{\circ}$ {(ejs)}. 
Four additional simulations have
%,jsres, 
Jupiter and Saturn with $a_{J}=5.43$~au and $a_{S}=7.30$~au,
$e_{J}=0.07$, $e_{S}=0.01$ and mutual inclination of
$0.2^{\circ}$ {(jres)} and the last four are performed with the same semi-major axis and
$e_{J}=0.07=e_{S}=0.03$ {(jresecc)}.

All simulations in set II \citep[see details in][]{2015Icar..252..161K} consist of  1000 planetesimals
and 100 planetary embryos extending between $0.5$ and $4.0$~pc.
%cjs
In 50 simulations, Jupiter and Saturn are on nearly circular ($e<0.01$)
orbits with the current radius {(cjsII)}, while in the other 50 runs %ejs 
Jupiter and Saturn are on orbits with small eccentricity ($e_J=e_S=0.1$) and with their observed semi-major axis {(ejsII)} . \\ 
In both set I and set II a time-step of 6 days has been used and each system
has been evolved for 200Myr. Embryos and planetesimals collide with each other while orbiting the Sun,
producing 3-4 rocky planets by the end of each simulation and leaving behind
some not accreted planetesimals. The resulting planetary systems are similar to our Solar system. 
We note, however, that none of the simulations, in either set, is able to correctly reproduce the Solar System properties
accurately; nevertheless those are among the best simulations available at the moment.
In every run, the collisions experienced
by each embryo are recorded and can be used to map the feeding zone
of each planet that form in any system.

\begin{center}
\begin{figure*}
\begin{centering}
\includegraphics[trim=0cm 0cm 0cm 1cm, clip=true, width=0.43\textwidth, clip=true]{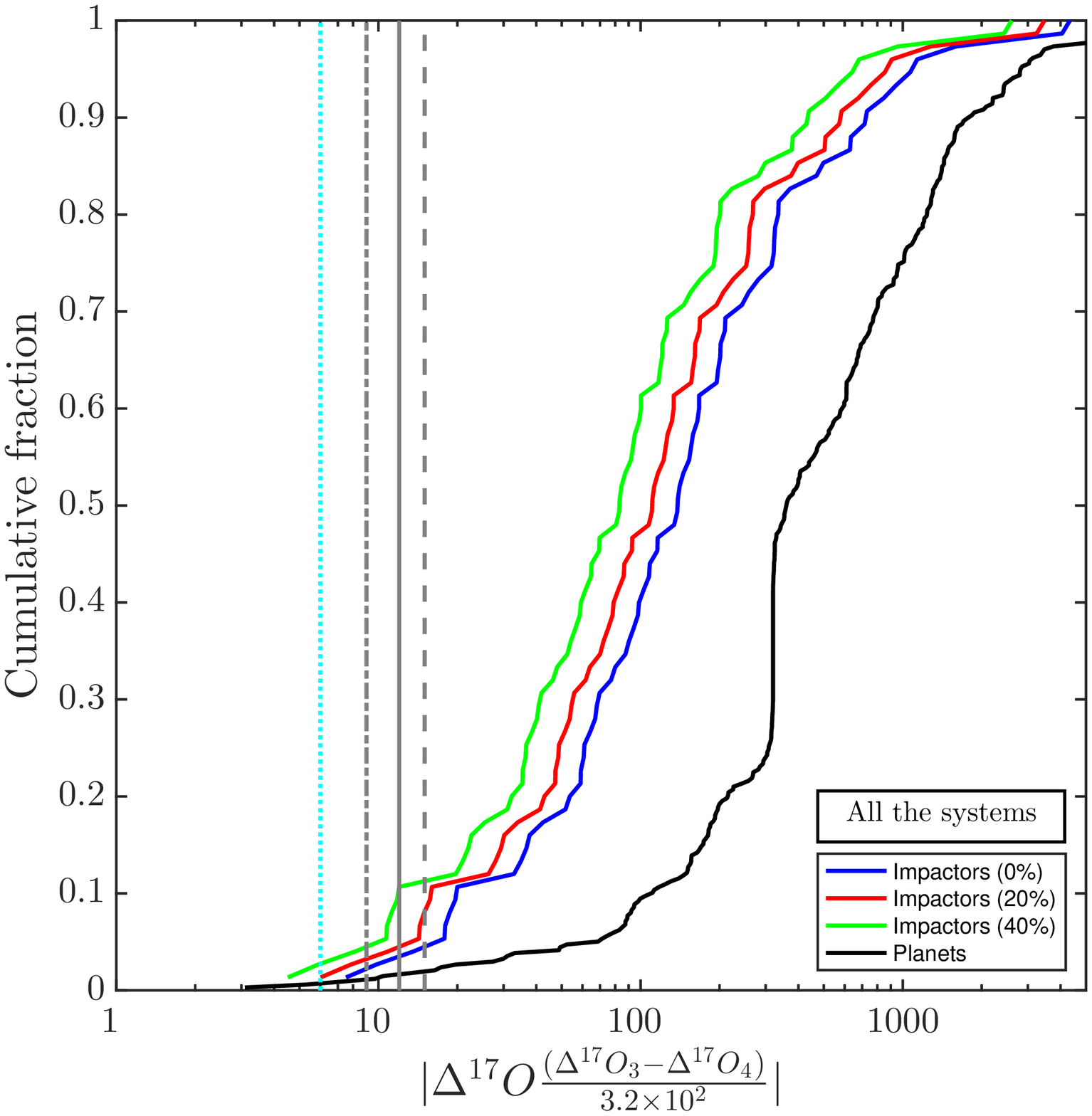}
\includegraphics[trim=0cm 0cm 0cm 1cm, width=0.43\textwidth, clip=true]{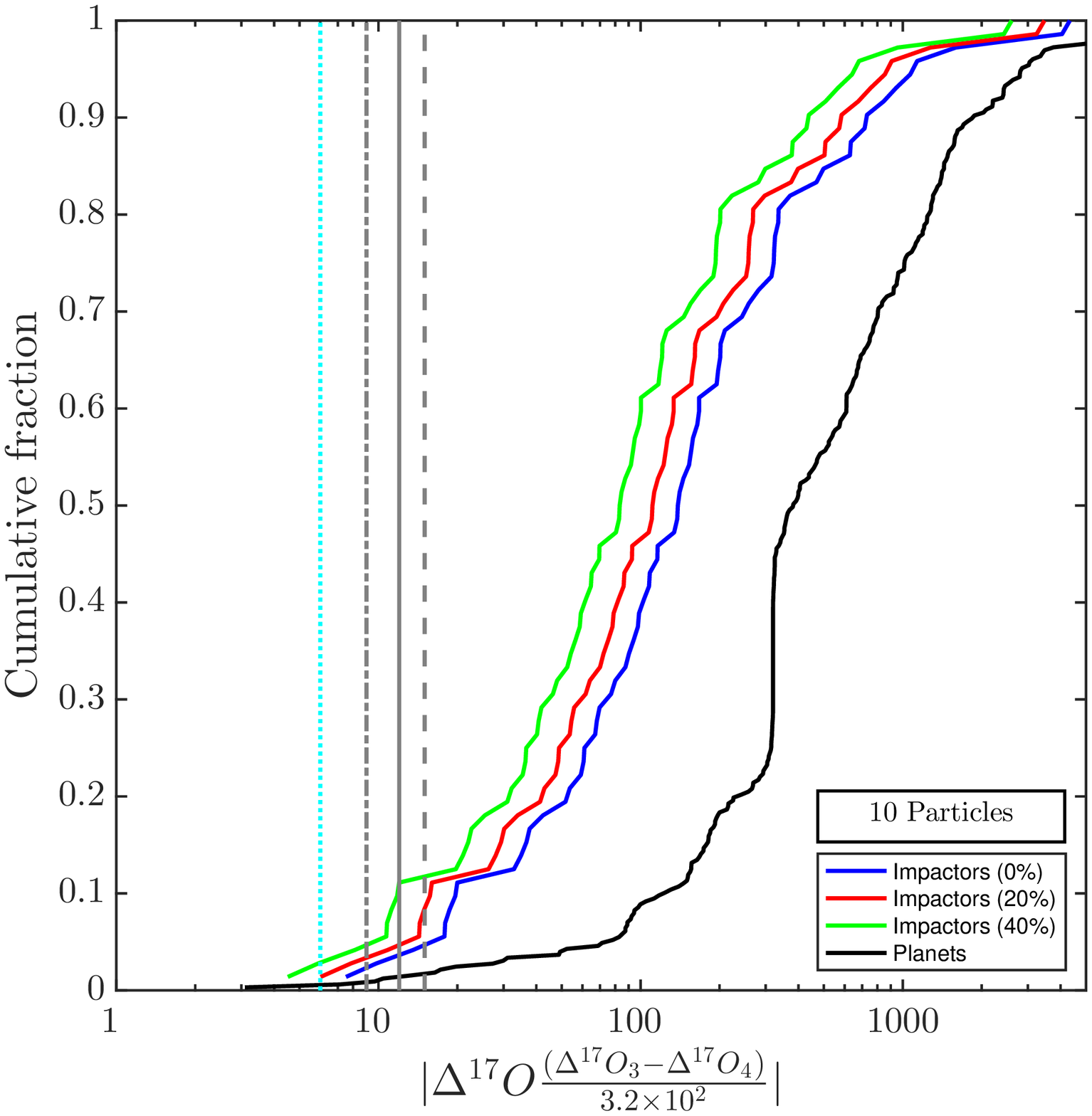}
\par\end{centering}
\centering{}\includegraphics[trim=0cm 0cm 0cm 1cm, width=0.43\textwidth, clip=true]{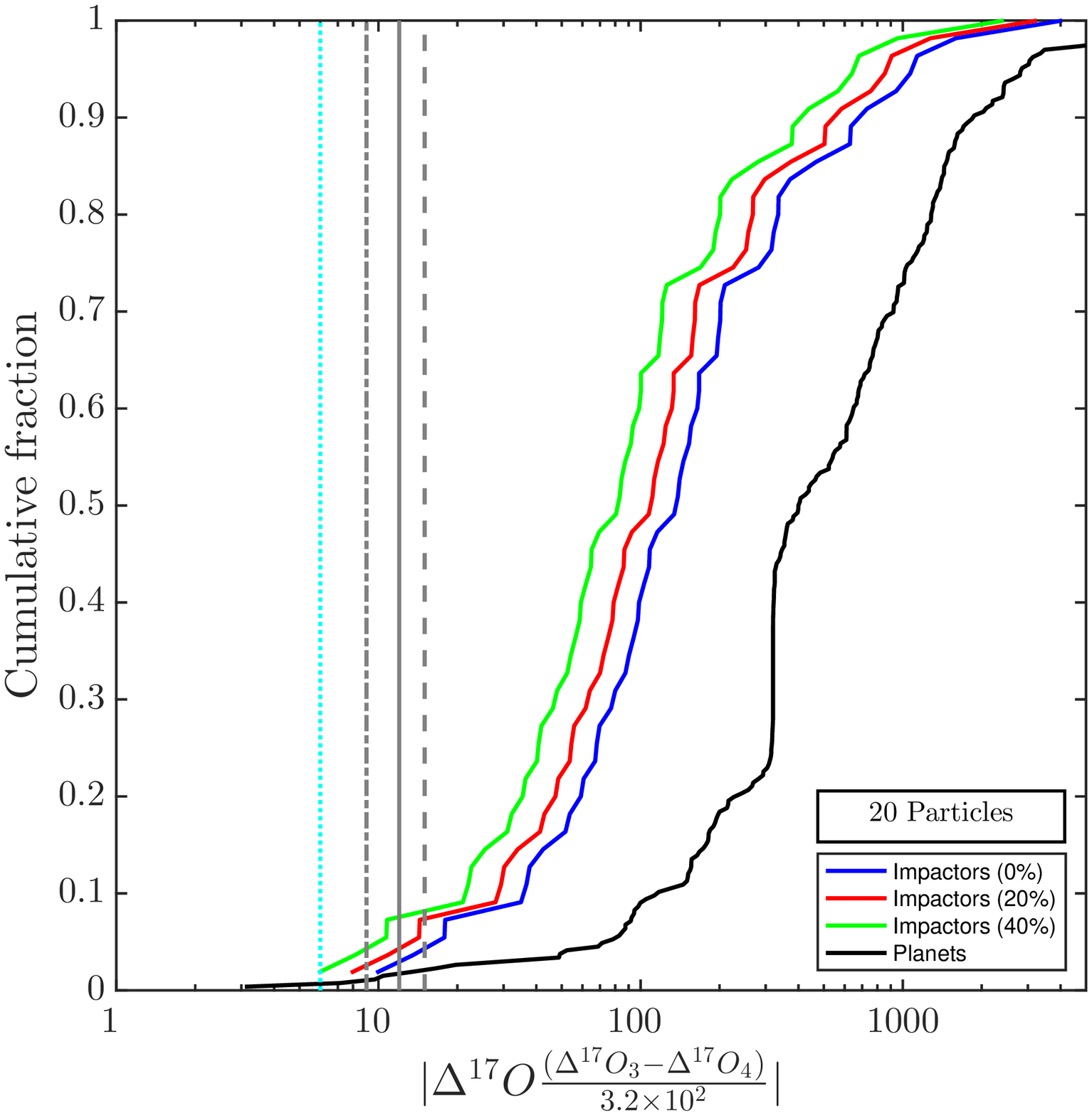}
\includegraphics[trim=0cm 0cm 0cm 1cm, width=0.43\textwidth, clip=true]{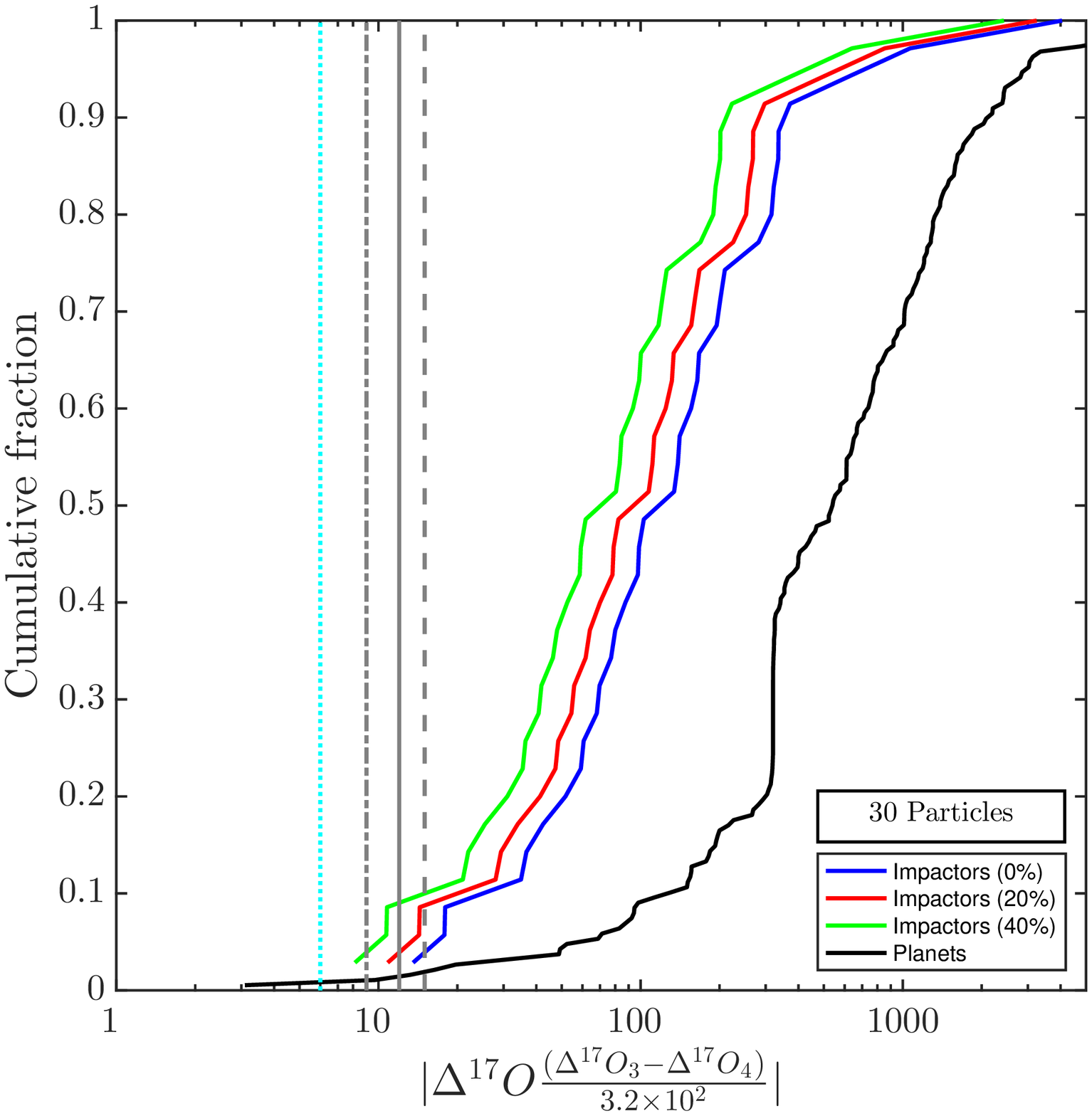}
\includegraphics[trim=0cm 0cm 0cm 1cm, width=0.43\textwidth, clip=true]{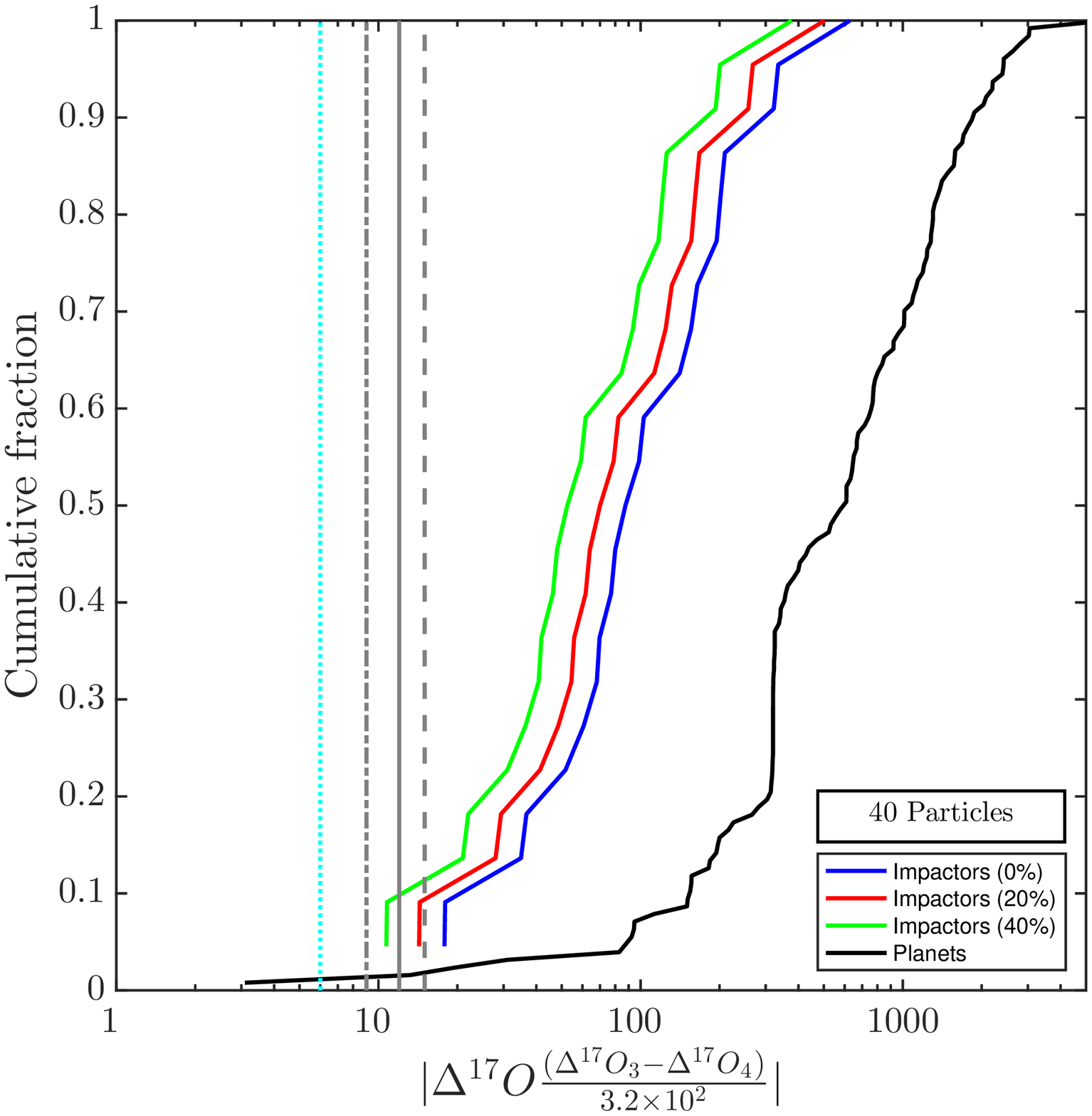}
\includegraphics[trim=0cm 0cm 0cm 1cm, width=0.43\textwidth, clip=true]{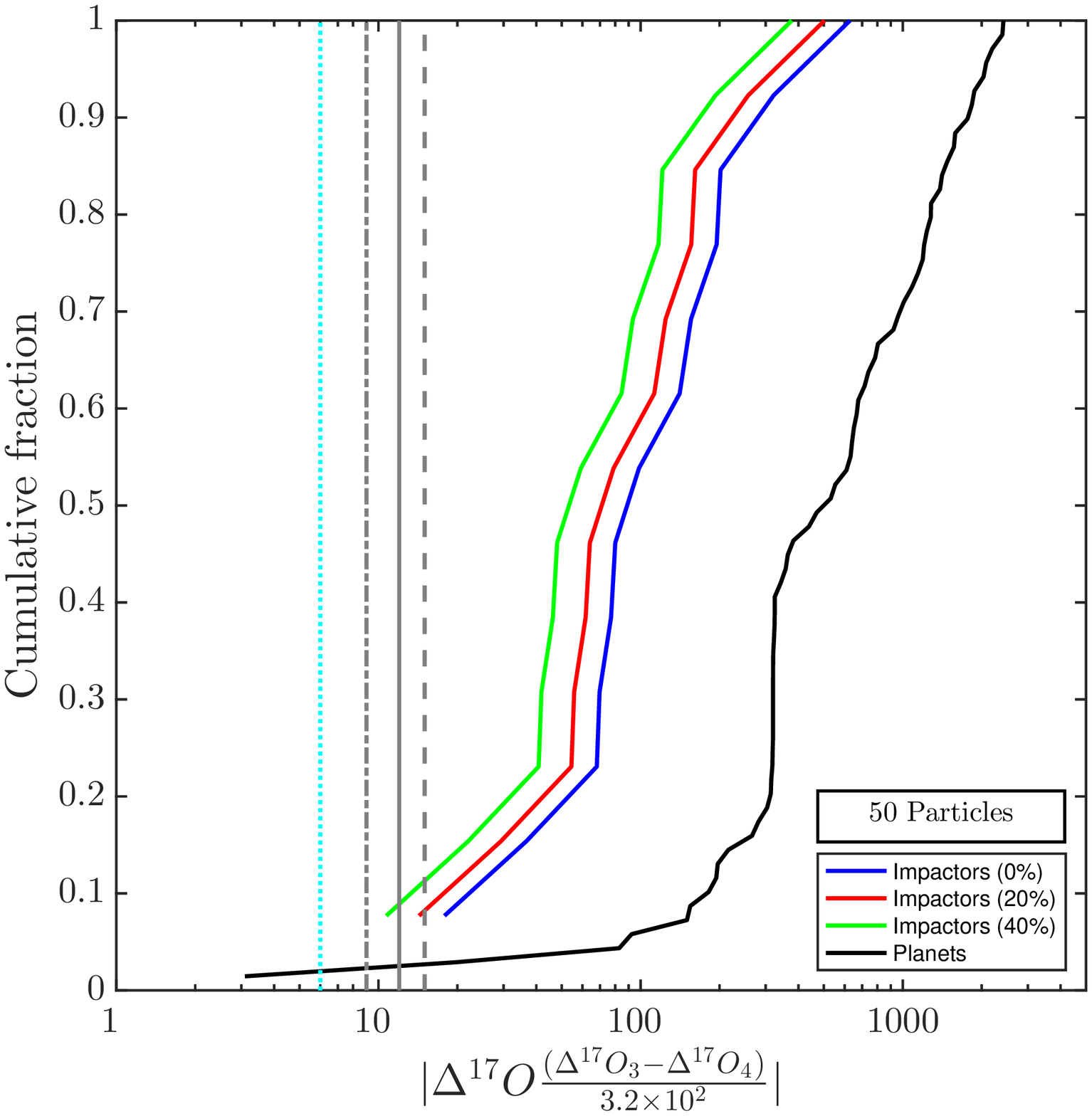}
\caption{The cumulative distribution of the absolute $\Delta^{17}O$ differences
between Earth analogs and Theia analogs, compared with the differences
between planets in the same system (black), obtained from the raw
simulations. Blue, red and green lines correspond to the cases of
0\%, 20\% and 40\% contribution of material from the planet to a
moon formed from these impacts, respectively. The vertical grey lines
depict the $\Delta^{17}O$ difference of the Earth-Moon system (dot-dashed and dashed
lines for $\pm\sigma$ around the mean value; central solid line).
The cyan dot vertical line depicts the more recent $1\sigma$ estimate
for the same difference \citep[see][]{2016Sci...351..493Y}. The same plot is shown for all
the systems (top left panel), regardless of the number of particles
that contributed to their formation, and planets and last impactors
composed of a minimum of 10, 20, 30, 40 
and 50 particles.}
\label{fig:The-cumulative-distribution}
\end{figure*}

\par\end{center}

\subsection{Earth and Mars analogs}\label{susect:analogs}
Following  \cite{2015Icar..252..161K, 2015Icar..258...14K} we identify an Earth analog as the
planet that have a final semi-major axis between $0.8$ and $1.2$~au and mass
$m>0.5M_{\oplus}$. Mars analog is defined as the next planet to
the Earth with $m>0.05M_{\oplus}$ while Theia analog is the last body
with $m>0.1M_{\oplus}$ that hit the Earth analog. 
These conditions
are fulfilled in 75  out of 140 simulations (22 from set I and 53 from set II).\\
The analog of the last Mars impactor is
the last planetary embryo that struck the planet and whose mass is larger than 0.01 times the mass of Mars.
This choice leaves us with 79 simulations useful for analysis (25 from set I and 54 from set II). \\
{ In order to maximise the number of simulations that can be analysed, we did not constrain Mars' semi-major axis.
However, if Mars analog is orbiting at $2$-$3$~au, it will likely be composed of material that had very distant initial orbits, smoothing the $\Delta^{17}O$  distribution and leading to more isotopically similar Earth and Theia. Thus we repeated our analysis imposing an upper limit of 2au for
Mars' semi-major axis. This reduces the number of simulations with proper analogs to 70, 19 from set I and 51 from set II (see also Appendix \ref{app:SMA}).}

\subsection{Composition calibration} \label{cal}
The oxygen isotope composition is often used as a tracer for Earth-Moon similarity because it is a good composition indicator and it 
has been measured for several bodies in the Solar System.
We used as a reference the value measured by \cite{HP14} ($\Delta^{17}O=12\pm3$~ppm) as done in Paper I.
Recently, \cite{2016Sci...351..493Y} using a different sample of rocks found a significantly smaller difference ($\Delta^{17}O=-1\pm5$~ppm).
However, using one of their samples they get a difference of $-16 \pm3$~ppm, even larger than \cite{HP14} estimate. 
Since the mismatch between the different estimates could be the result of the choice of the rock samples and not of an intrinsic property
of the Moon, we decided to compare our results with both available estimates.

In order to calculate the $\Delta^{17}O$ ($\equiv\delta^17O-0.52\delta^16O$) of each planet we used the same procedure as
described in Paper I and in \cite{PS07}. We assigned a $\Delta^{17}O$
value to each body, according to its initial position
in the proto-planetary disk. { Since the initial $\Delta^{17}O$ distribution in the proto-planetary disk is in principle unknown we assume a simple } linear gradient of $\Delta^{17}O$
with the distance from the Sun 

\begin{equation}
\Delta^{17}O(r)=c_{1}r+c_{2}
\label{eqcal}
\end{equation}

where the two free parameters can be calibrated using the known Earth
and Mars $\Delta^{17}O$ values ($0$ppm and $320$ppm respectively, \citealt{Fr99}). The final composition of each planet is the mass weighted average of
the composition of all the planetesimals and embryos that contributed to its formation.
%To take into account the stochasticity of the process, we evaluated the error as the Poissonian $\sigma$ on the composition
%obtained. 
We compared the composition of each last impactor on the Earth to that observed for the Earth-Moon system and we obtained the  
fraction of compatible Earth-Theia analogs pairs. 
{ Several different calibrations have been used in literature \citep[see e.g.][]{2015Icar..252..161K, 2016Sci...351..493Y}.  Following the recent results described in \cite{Dau17}, we tested the giant impact scenario using
a sharp contrast between the inner and outer proto-planetary disk, modelled using a step function.
The results of this analysis are shown in Appendix \ref{app:SF}.}

\subsection{Statistical analysis: bootstrapping vs granularity} 

While the number of simulated embryos can be considered
realistic, real proto-planetary disks are composed by orders of magnitude larger number of planetesimals than what simulated
\citep[see][]{2015Icar..258...14K},
possibly leading to a poorly realistic representation of its impact history.
To estimate the effects of low resolution, we considered either all the systems or only the ones whose analogs are composed from a 
minimum number of particles (10 to 50). 
To take into account the stochasticity of the process
caused by the low statistics, we evaluated the uncertainty as the Poissonian $\sigma$ on the fraction of compatible systems as the square root of the number of positive events (compatible samples), divided by the total number of events (number of simulations with analogs).

\begin{center}
\begin{figure*}
\begin{centering}
\includegraphics[width=0.5\textwidth]{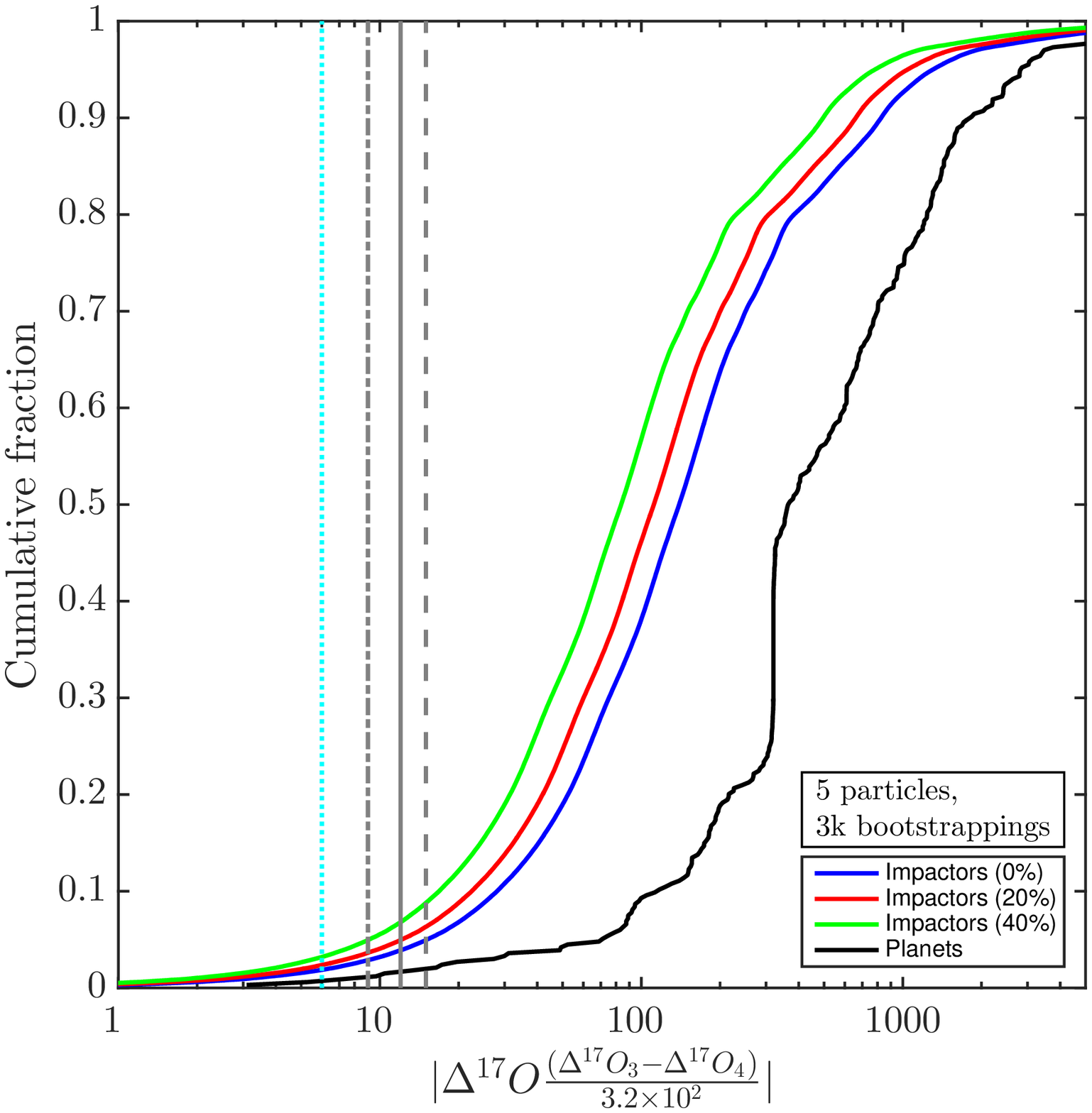}\includegraphics[width=0.5\textwidth]{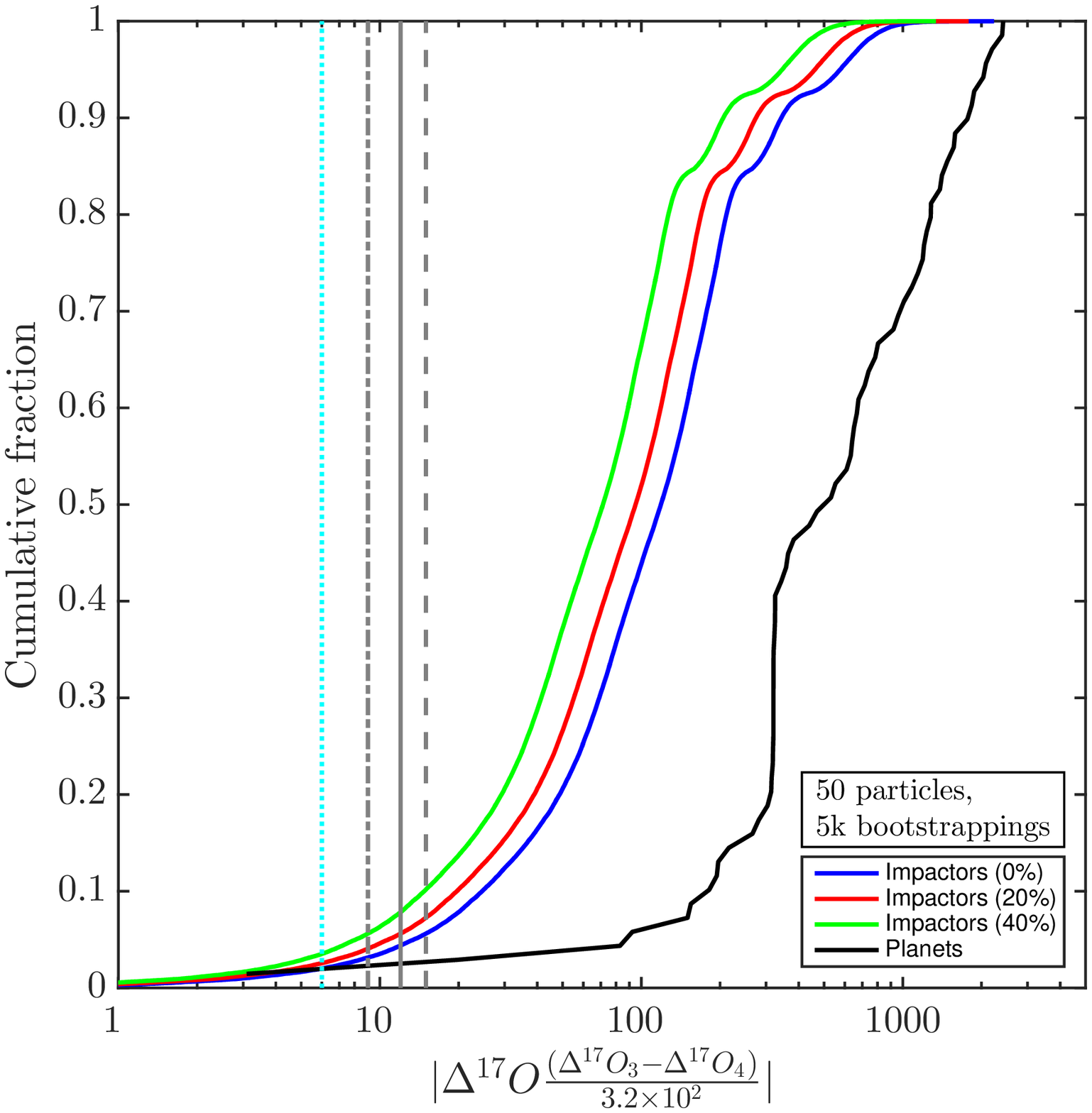}
\par\end{centering}

\caption{The same as in Figure \ref{fig:The-cumulative-distribution}, but
for the bootstrapped simulations. The cumulative distributions of $\Delta^{17}O$ are shown for Earth
and Theia analogs composed by at least 5 particles and $3000$ bootstrapped samples (left panel) and for a threshold of 50 
particles and $5000$ bootstrapped samples (right panel).}
\label{cum_boots}
\end{figure*}

\par\end{center}

\begin{center}
\begin{figure}
\begin{centering}
\includegraphics[clip=true,width=0.5\textwidth]{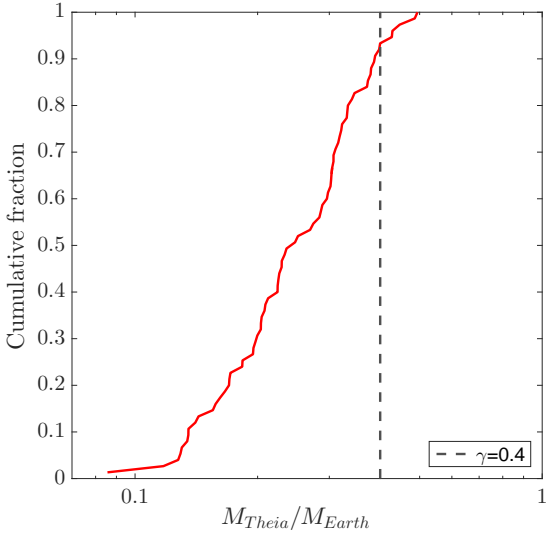}
\par\end{centering}

\centering{}\caption{The final mass ratio between each Theia analog and the relative Earth analog ($\gamma$, red solid
line).  The vertical grey line represents the $\gamma=0.4$ threshold.}
\label{fig:mass_ratio}
\end{figure}

\par\end{center}

\begin{table*}
\caption{The mean fraction (and relative $1\sigma$ Poissonian error) of compatible Earth-Theia 
analogs is given for different minimum numbers of particles
composing the analogs ($N_{\text{min}})$, for different limits on the observed $\Delta^{17}O$ (6, 9, 12 and 15ppm) and mixing fractions (0\%, 20\%, 40\%).  
$N_{\text{cases}}$ is the number of simulations with proper analogs. { Results obtained imposing an upper limit of $2$~au on the semi-major axis of Mars' analogs are reported in brackets (see Appendix \protect\ref{app:SMA} for more information).}}
\begin{tabular}{ccccccccc}
\hline
\hline
\multirow{2}{*}{$N_{\text{min}}$} & \multirow{1}{*}{$\Delta^{17}O$} & \multicolumn{2}{c}{$0%
$\%} & \multicolumn{2}{c}{2$0%
$\%} & \multicolumn{2}{c}{4$0%
$\%} & \multirow{2}{*}{$N_{\text{cases}}$}\tabularnewline
 & ppm & Mean (\%) & $\pm\sigma$ (\%) & Mean (\%) & $\pm\sigma$(\%) & Mean (\%) & $\pm\sigma$(\%) & \tabularnewline
\hline 
\multirow{4}{*}{0} & 6 & $0$ & $0$ & $0$ & $0$ & $2.7 (1.4)$ & $1.9 (1.4)$ & \multirow{4}{*}{75 (70)}\tabularnewline
 & 9 & $1.3 (1.4)$ & $1.3 (1.4)$ & $2.7 (1.4)$ & $1.9 (1.4)$ & $4.0 (2.9)$ & $2.3 (2.0)$ & \tabularnewline
 & 12 & $2.7  (1.4)$ & $1.9 (1.4)$ & $4.0 (2.9)$ & $2.3 (2.0)$ & $10.7 (10.0)$ & $3.8 (3.8)$ & \tabularnewline
 & 15 & $4.0 (2.9)$ & $2.3 (2.0)$ & $8.0 (7.1)$ & $3.3 (3.2)$ & $10.7 (10)$ & $3.8 (3.8)$ & \tabularnewline
\tabularnewline
\multirow{4}{*}{10} & 6 & $0$ & $0$ & $0$ & $0$ & $2.8 (1.5)$ & $2.0 (1.5)$ & \multirow{4}{*}{72 (67)}\tabularnewline
 & 9 & $1.4 (1.5)$ & $1.4 (1.5)$ & $2.7 (1.5)$ & $1.9 (1.5)$ & $4.1(3.0)$ & $2.4 (2.1)$ & \tabularnewline
 & 12 & $2.8 (1.5)$ & $2.0 (1.5)$ & $4.2 (2.9)$ & $2.4 (2.1)$ & $11.1 (10.4)$ & $3.9 (3.9)$ & \tabularnewline
 & 15 & $4.2 (3.0)$ & $2.4 (2.1)$ & $8.3 (7.5)$ & $3.4 (3.3)$ & $11.1 (10.4)$ & $3.9 (3.9)$ & \tabularnewline
\tabularnewline
\multirow{4}{*}{20} & 6 & $0$ & $0$ & $0$ & $0$ & $1.8 (0)$ & $1.8 (0)$ & \multirow{4}{*}{55 (50)}\tabularnewline
 & 9 & $0$ & $0$ & $1.8$ & $1.8$ & $3.6 (2.0)$ & $2.6 (2.0)$ & \tabularnewline
 & 12 & $1.8 (0)$ & $1.8 (0)$ & $3.6 (2.0)$ & $2.6 (2.0)$ & $7.3 (6.0)$ & $3.6 (3.5)$ & \tabularnewline
 & 15 & $3.6 (2.0)$ & $2.6 (2.0)$ & $7.3 (6.0)$ & $3.6 (3.5)$ & $7.3 (6.0)$ & $3.6 (3.5)$ & \tabularnewline
\tabularnewline                                                                                                                                                                                                                                                                                                                                                                                                                                                                                                                                                                                                                                                                                                                                                                                                                                                                                                                                                                                                                                                                                                                                                                                                                    
\multirow{4}{*}{30} & 6 & $0$ & $0$ & $0$ & $0$ & $0 (3.1)$ & $0 (3.1)$ & \multirow{4}{*}{35 (32)}\tabularnewline
 & 9 & $0$ & $0$ & $0$ & $0$ & $2.9 (3.1)$ & $2.9 (3.1)$ & \tabularnewline
 & 12 & $0$ & $0$ & $2.9 (3.1)$ & $2.9 (3.1)$ & $8.6 (9.4)$ & $4.9 (5.4)$ & \tabularnewline
 & 15 & $2.9 (3.1)$ & $2.9 (3.1)$ & $8.6 (9.3)$ & $4.9 (5.4)$ & $8.6 (9.4)$ & $4.9 (6.7)$ & \tabularnewline
\tabularnewline
\multirow{4}{*}{40} & 6 & $0$ & $0$ & $0$ & $0$ & $0$ & $0$ & \multirow{4}{*}{22 (21)}\tabularnewline
 & 9 & $0$ & $0$ & $0$ & $0$ & $0$ & $0$ & \tabularnewline
 & 12 & $0$ & $0$ & $0$ & $0$ & $7.7 (9.5)$ & $5.4 (5.4)$ & \tabularnewline
 & 15 & $0$ & $0$ & $9.1 (11.8)$ & $6.4 (8.3)$ & $9.1 (11.8)$ & $6.4 (8.3)$ & \tabularnewline
\tabularnewline
\multirow{4}{*}{50} & 6 & $0$ & $0$ & $0$ & $0$ & $0$ & $0$ & \multirow{4}{*}{13 (12)}\tabularnewline
 & 9 & $0$ & $0$ & $0$ & $0$ & $0$ & $0$ & \tabularnewline
 & 12 & $0$ & $0$ & $0$ & $0$ & $7.7 (8.3)$ & $7.7 (8.3)$ & \tabularnewline
 & 15 & $0$ & $0$ & $7.7 (8.3)$ & $7.7 (8.3)$ & $7.7 (8.3)$ & $7.7 (8.3)$ & \tabularnewline
\hline 
\end{tabular}
\label{tab1}
\end{table*}

To further explore and quantify the error due to granularity we used a bootstrap technique with replacement. 
We took the original distribution of the initial semi-major axis of the planetesimals composing each Earth, { Mars} and Theia analog and we
extracted a new, equally populated, distribution of semi-major axis from it   \citep[see also][]{2015Icar..258...14K}.
Only systems which analogs are composed of more 
than 5 particles ($73$ simulations) have been taken into consideration. We repeated the bootstrapping 5000
times for each distribution when the set of possible bootstrap samples was rich enough (i.e. more than minimum 10 particles, 
while we bootstrapped only 3000 times for bodies 
formed by at least 5 particles).\footnote{The reason of this choice stands in the fact that the number of the possible bootstrapped combinations
is only 210 for bodies composed by 4 particles and 3024 in the case of 5 planetesimal.} \\
The Poissonian error on the fractions has been evaluated in a similar way as done in the case of the raw simulations. 
We show here an example of the error calculation, in which we assume that we have N simulations
among which two give non null fractions,  $f_1$ and $f_2$, in their bootstrapped sample.
In this case, the total fraction obtained considering all the simulations is given by the sum of the probabilities of the possible outcomes
\begin{equation}
P=P2+P1=[2f_1f_2+f_1(1-f_2)+f_2(1-f_1)]/N=(f_1+f_2)/N
\label{eq:err}
\end{equation}
The total error can be evaluated applying the error propagation on Equation \ref{eq:err} and
assuming $\sigma(P1)=\sqrt{1}[f_1(1-f_2)+f_2(1-f_1)]$ and $\sigma(P2)=\sqrt{2}f_1f_2$.
We evaluated the Poissonian error following a generalised version of this method, and taking into account
only the bootstrapped samples that result in a non negligible fraction $f_i$.

\section{Results}\label{sec:results}

\begin{table*}
\caption{The same as in Table \ref{tab1} but for systems with minimum $5$ ($73$ simulations) or $50$ ($13$ simulations) particles composing each analog and respectively
3000 or 5000 bootstrappings. { Results obtained imposing an upper limit of $2$~au on the semi-major axis of Mars' analogs are reported in brackets (see Appendix \protect\ref{app:SMA} for more information).}}
\begin{tabular}{cccccccc}
\hline
\hline
\multirow{1}{*}{$\Delta^{17}O$} & \multirow{2}{*}{$N_{\text{min}}$, $N_{\text{bss}}$} & \multicolumn{2}{c}{$0%
$\%} & \multicolumn{2}{c}{2$0%
$\%} & \multicolumn{2}{c}{4$0%
$\%}\tabularnewline
ppm &  & Mean (\%) & $\pm\sigma$ (\%) & Mean (\%) & $\pm\sigma$(\%) & Mean (\%) & $\pm\sigma$(\%)\tabularnewline
\hline 
\multirow{2}{*}{$6$} & $5$, $3$k & $1.9 (1.6)$ & $1.4 (1.2)$ & $2.4 (2.0)$ & $1.6 (1.4)$ & $3.2 (2.8)$ & $2.0 (1.8)
$\tabularnewline %73
 & $50$, $5$k & $2.0 (2.2)$ & $1.9 (1.9)$ & $2.6 (2.8)$ & $2.4 (2.5)$ & $3.5 (3.8)$ & $3.2 (3.4)$\tabularnewline %13
\multirow{2}{*}{$9$} & $5$, $3$k & $2.9 (2.5)$ & $1.8 (1.6)$ & $3.6 (3.2)$ & $2.1 (2.0)$ & $4.9 (4.4)$ & $2.5 (2.4)$\tabularnewline %73
 & $50$, $5$k & $3.1 (3.4)$ & $2.9 (3.1)$ & $4.0 (4.4)$ & $3.7 (3.9)$ & $5.6 (6.1)$ & $4.9 (5.3)$\tabularnewline %13
\multirow{2}{*}{12} & $5$, $3$k & $3.9 (3.4)$ & $2.2 (2.1)$ & $4.9 (4.4)$ & $2.5 (2.4)$ & $6.8 (6.1)$ & $3.0 (2.9)$\tabularnewline %73 68
 & $50$, $5$k & $4.4 (4.7)$ & $3.9 (4.2)$ & $5.6 (6.1)$ & $4.9 (5.3)$ & $7.8 (8.5)$ & $6.5 (7.0)$\tabularnewline %13 12
\multirow{2}{*}{15} & $5$, $3$k & $4.9 (4.4)$ & $2.5 (2.4)$ & $6.3 (5.6)$ & $2.9 (2.8)$ & $8.8 (8.0)$ & $3.4 (3.4)$\tabularnewline %68
 & $50$, $5$k & $5.6 (6.1)$ & $4.9 (5.3)$ & $7.3 (7.9)$ & $6.1 (6.6)$ & $10.2 (11.0)$ & $8.0 (8.6)$\tabularnewline %12
\hline 
\end{tabular}
\label{tab2}
\end{table*}

\subsection{The composition of the Moon} 
When adopting an upper limit of $15$ppm for the  composition difference, we find the probability of having an impact between an Earth and a Theia analog
with similar oxygen isotope composition to be between 4\% and 10.7\%, depending on the percentage of 
mixing allowed (from 0\% to 40\%).  When considering the Poissonian $1\sigma$ error the  probability is $<14.5$\%. 
Another study \citep{2016Sci...351..493Y} recently suggested a smaller difference of at most $6$ppm. When taking this maximal difference for oxygen composition
we get 0\%-2.7\% probability of compatibility. Including the Poissonian error 
the same probability is
between 0\% and 4.6\%. These percentages change when considering thresholds for the 
minimum number of bodies in each analog (see Figures \ref{fig:The-cumulative-distribution}  and Table \ref{tab1}).

Our results are also affected by observational errors that can be taken into account comparing the composition of the simulated systems 
to the lower limit, mean and upper limit of the corresponding observed quantity.
Table \ref{tab1} shows the fractions of compatible planet-impactor
systems depending on the given constraint on the observed Earth-Moon $\Delta^{17}O$ difference.

Figure \ref{cum_boots} shows
the cumulative distribution of the differences for a minimum of 5 particles composing each body
and 3000 bootstrapped combinations (left panel) for a threshold of 50 particles contributing to analog and 5000
combinations (right panel). 
Bootstrapping 3000 times on all the systems formed by at least 5 particles,  
allowing for mixing and considering the $1\sigma$ error, the probability of compatibility is  
smaller than 12.2\% (5.2\% for $6$ppm). 
Adopting a threshold of 50 particles and bootstrapping 5000 times on each distribution 
the probability is 5.6\%-18.2\% (2\%-6.7\% for $6$ppm).  
The fractions obtained for different thresholds, number of bootstrappings and 
observational constraints are listed in Table \ref{tab2}.
 { All the results slightly vary (without a clear trend) when setting  an upper limit ($2$~au) for Mars' analogs semi-major axis (see Table \ref{tab1}, Appendix \ref {app:SMA} and Figures \ref{fig:cum_2au}  and \ref{fig:cum_boots_mars2au} for more details.)}
Both Figure \ref{fig:The-cumulative-distribution} and \ref{cum_boots} {(as well as Figures \ref{fig:cum_2au} and \ref{fig:cum_boots_mars2au})} show that 
the Earth-Theia analog couples are sistematically more similar in compositions compared to the
other planets.
{ Among all the initial configurations used in the simulations the ones with Jupiter and Saturn on eccentric orbits, in particular ejsII, produce more compatible Earth-Theia couples. The configuration with slightly eccentric  orbits for Jupiter and Saturn (cjsecc) is the second most productive. Set II produces a larger fraction of positive cases than set I.}

It was shown that more mass can be extracted from the proto-Earth and mix into the Moon following the impact, if the impactor mass is comparable to that of the proto-Earth \citep{Ca12}.
In order to get about a half of the material in the proto-lunar disk
coming from the proto-Earth, the mass ratio between
Theia and the Earth ($\gamma$) must be larger than $0.4$ \citep[see][]{Ca12}.\\
We evaluated the cumulative distribution of the mass ratios for all the simulations with proper analogs (see Figure \ref{fig:mass_ratio})
and out of 75 simulations with analogs, only 5 have $\gamma>0.4$. This
corresponds to a 6.7 \% { (6.5\% when considering an upper limit for Mars' semi-major axis)} probability that such an event could happen.
Even if this probability is comparable to the one obtained for similar composition impacts, 
the process is efficient only when the impact involves extremely high angular momentum for the Earth-Moon system.
{ As shown by \cite{Ruf17}, this condition is not easily attained, because of angular momentum drain during the impact.} 
Moreover, the high angular momentum needs to be dissipated later on, which requires evection resonance due to the Sun to work efficiently \citep{Cu12}.

\begin{figure*}
\centering
\includegraphics[trim=4cm 0cm 4.5cm 1cm, clip=true, width=0.45\textwidth]{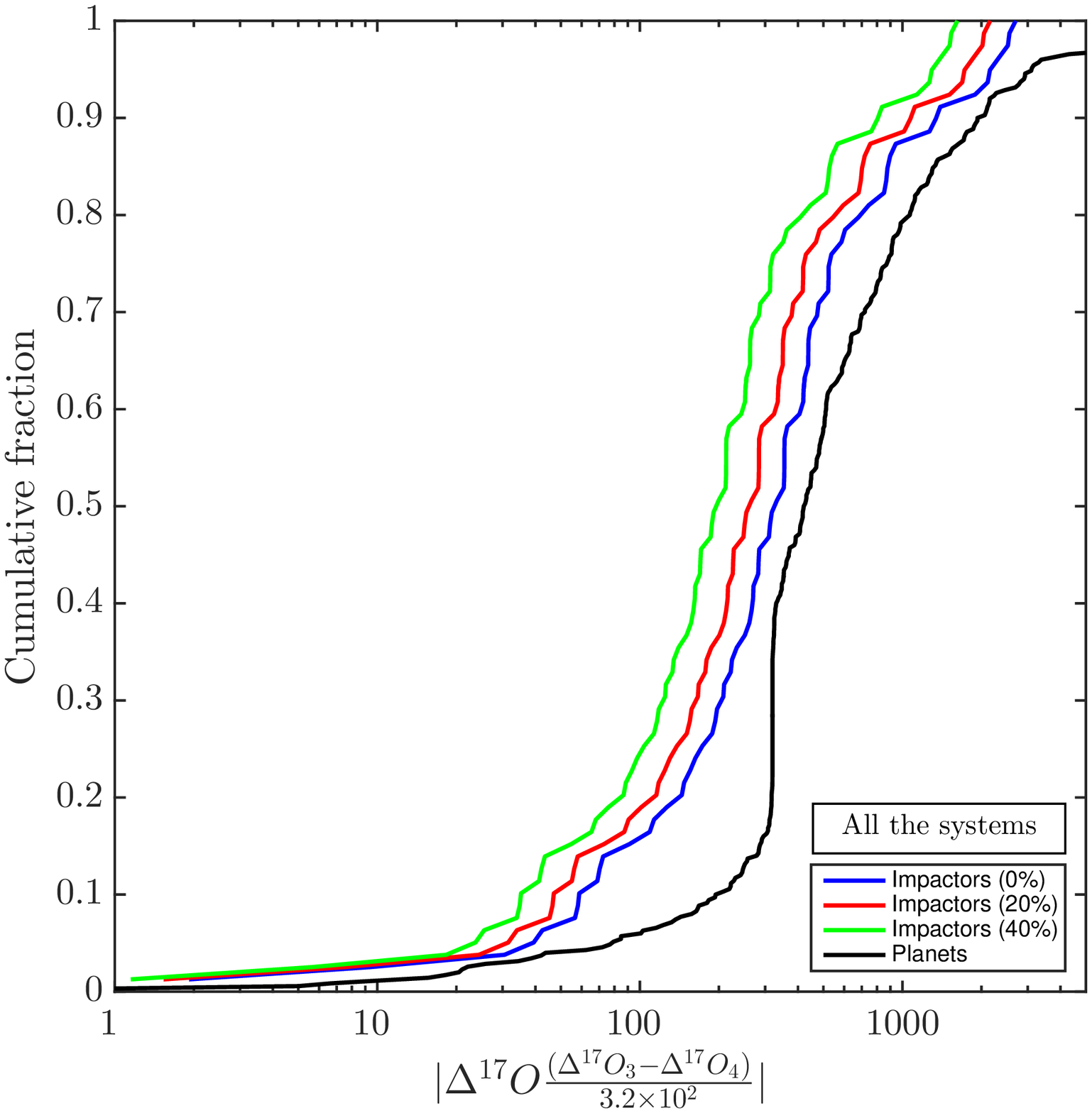}\includegraphics[trim=4cm 0cm 4.5cm 1cm,clip=true,width=0.45\textwidth]{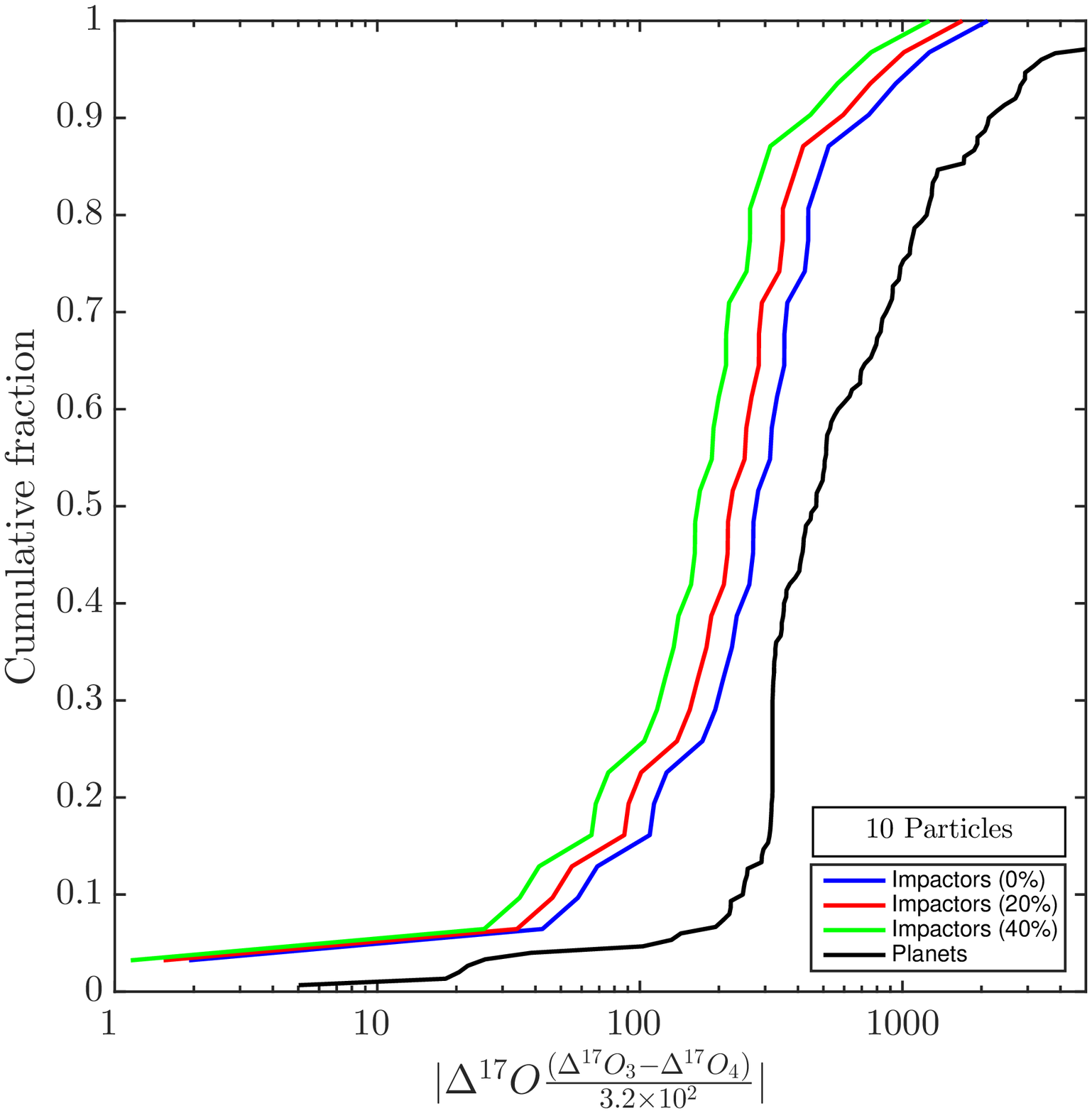}
\includegraphics[trim=4cm 0cm 4.5cm 1cm,clip=true, width=0.45\textwidth]{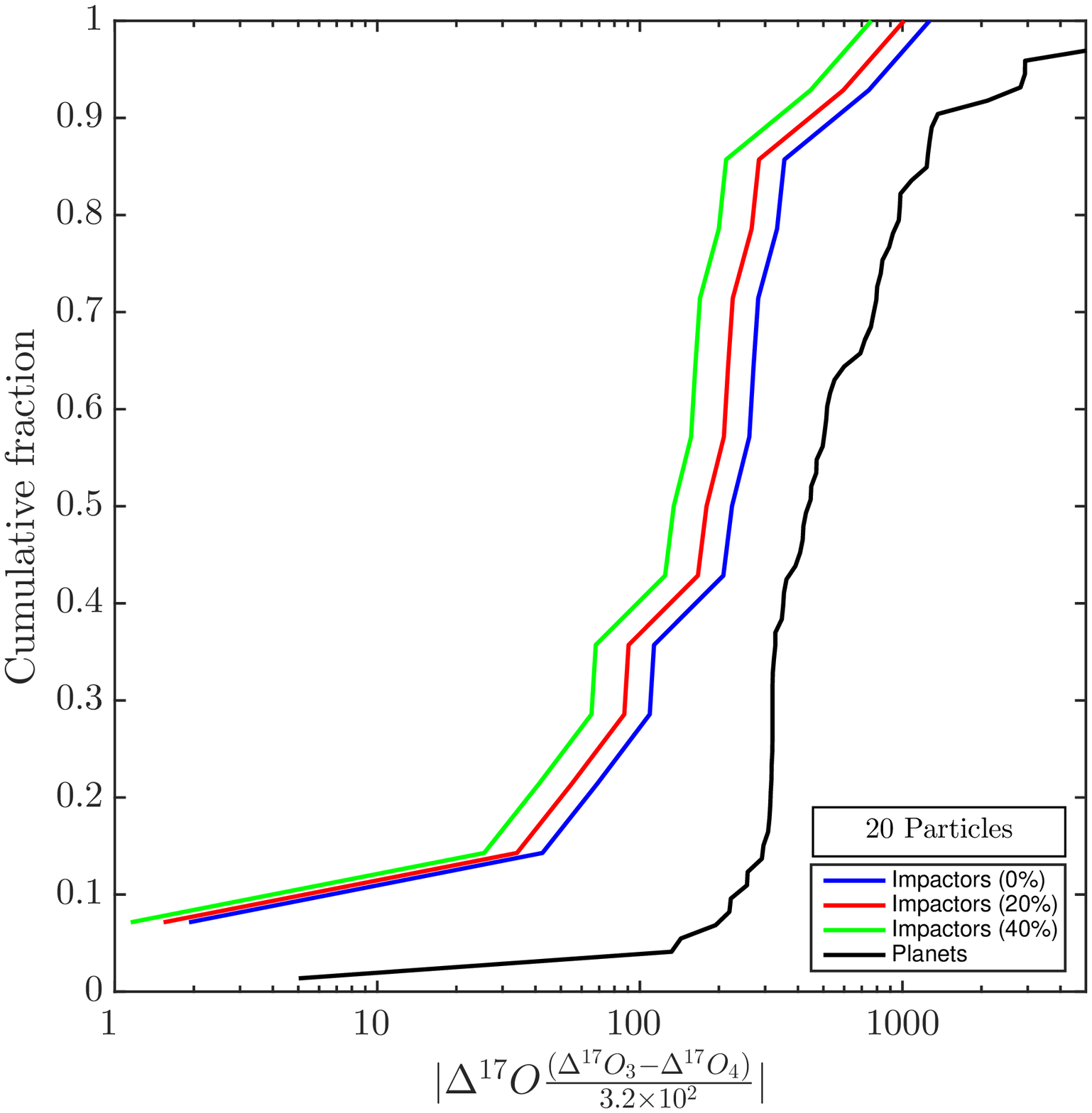}\includegraphics[trim=4cm 0cm 4.5cm 1cm,clip=true,  width=0.45\textwidth]{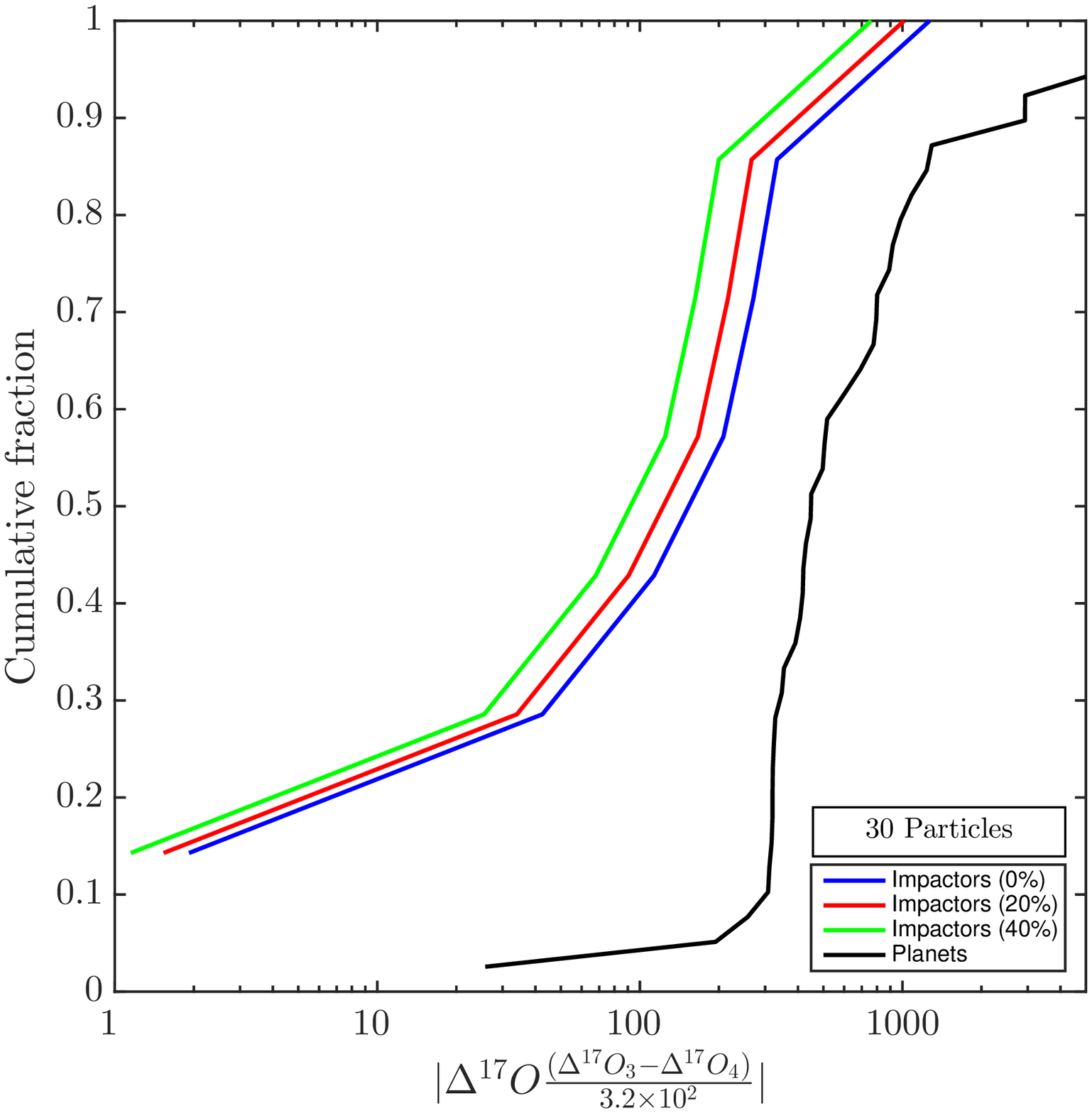}
\includegraphics[trim=4cm 0cm 4.5cm 1cm,clip=true,  width=0.45\textwidth]{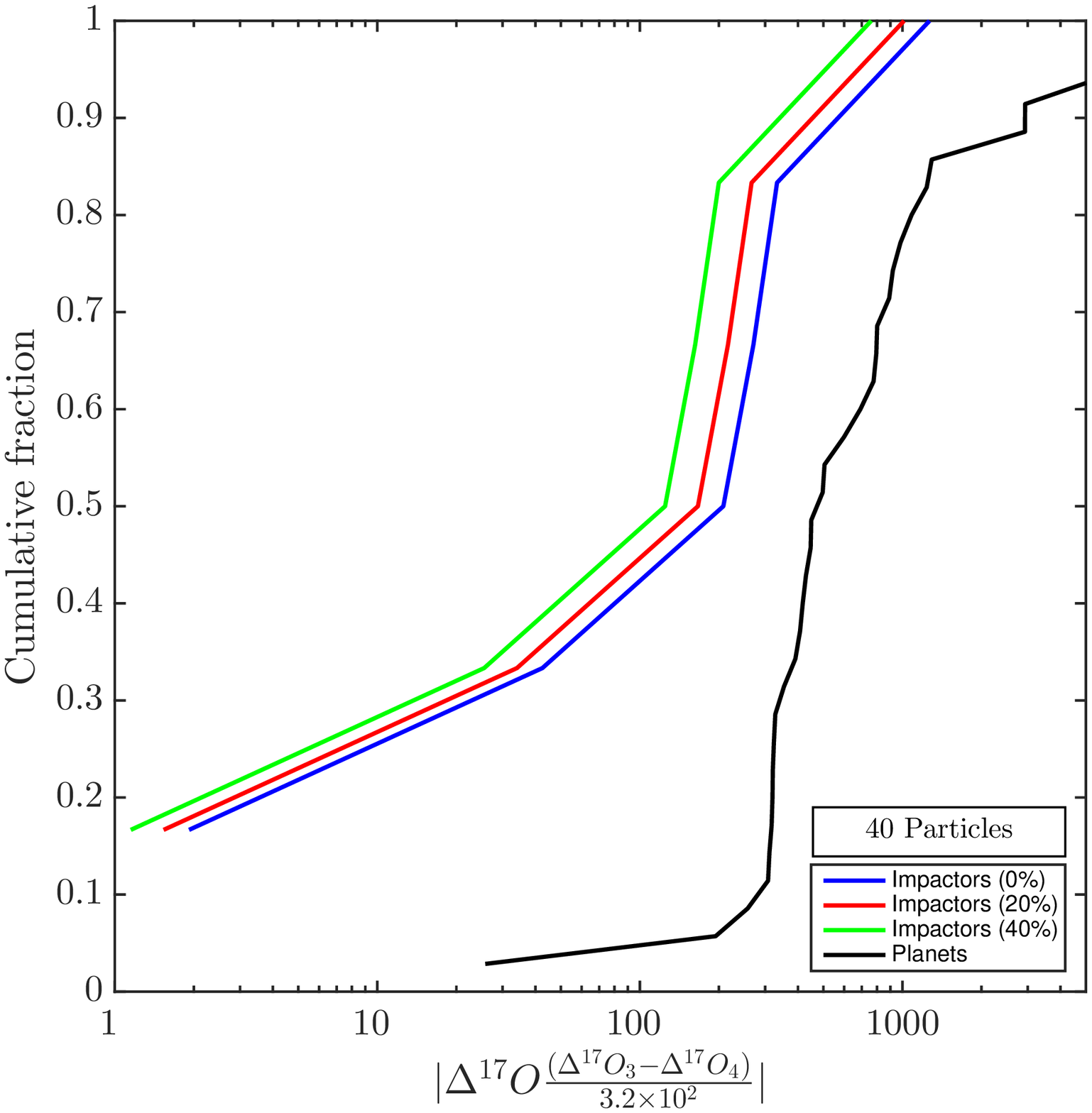}\includegraphics[trim=4cm 0cm 4.5cm 1cm,clip=true,  width=0.45\textwidth]{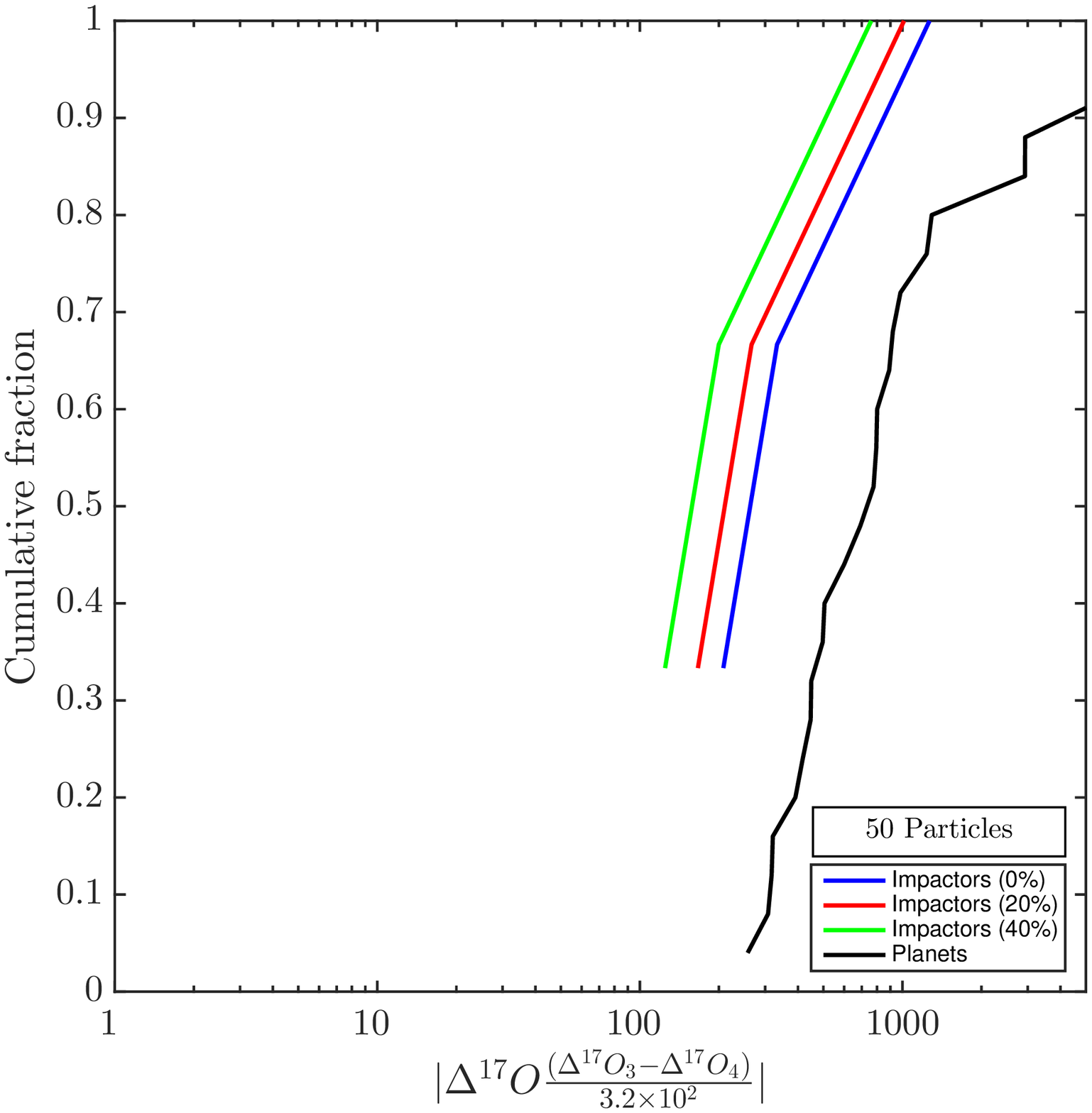}
\caption{The same as Figure \ref{fig:The-cumulative-distribution} but for Mars'
and its last impactor analogs. Recent SPH simulation \protect\cite[see][and private communication]{Cit15} showed that large mixing fractions are
possible. {The contribution of the impactor can be up to 70\% of the total mass of the proto-lunar disk from which Phobos and Deimos form. However, to be conservative, we adopted the same percentages used for the Earth-Moon system.} The difference between planets and planet-impactor couples increases with the mixing fraction.}
\label{fig:mars_cum}
\end{figure*}

\begin{figure*}
\begin{centering}
\includegraphics[width=0.5\textwidth]{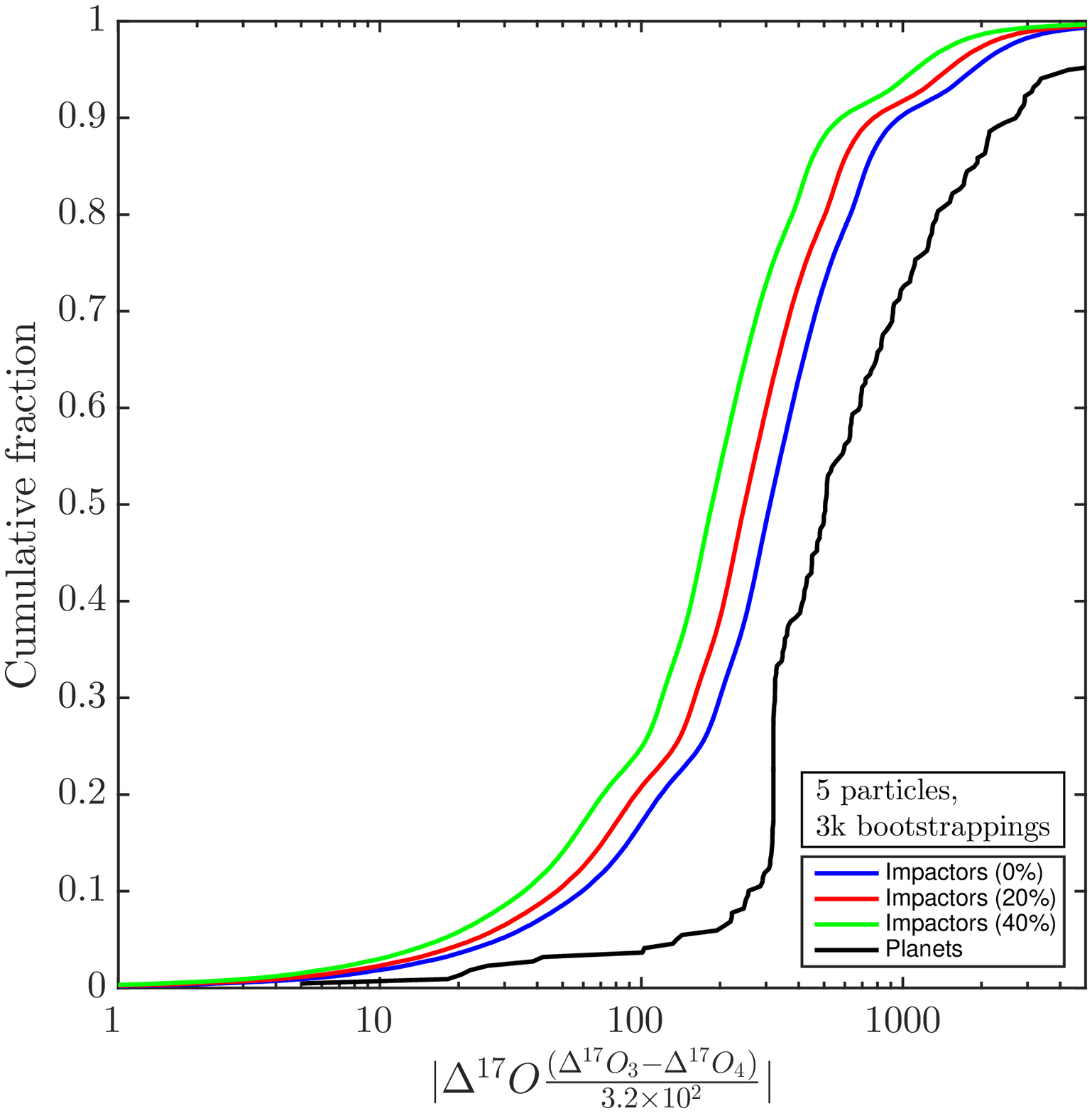}\includegraphics[clip=true,width=0.5\textwidth]{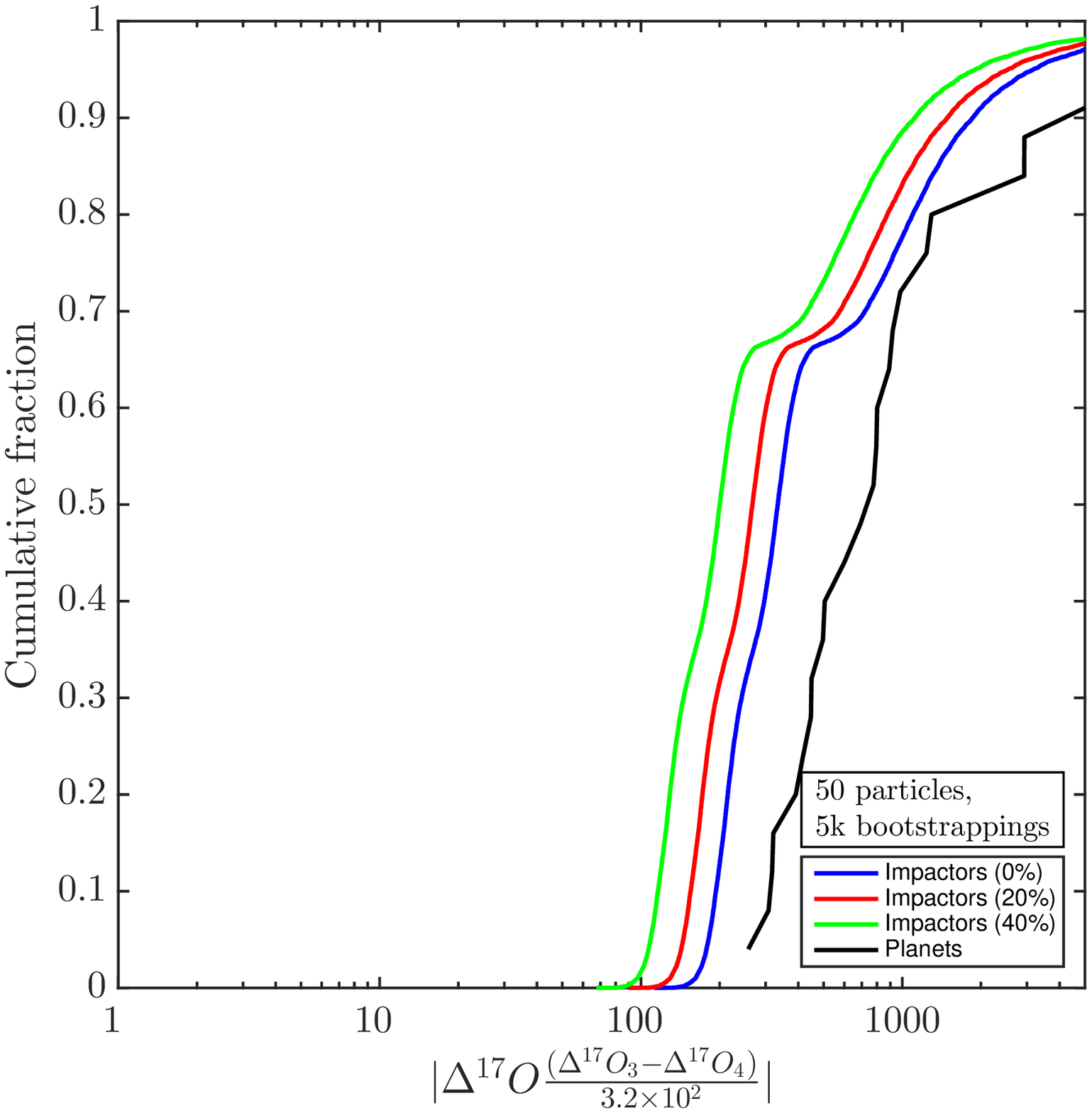}
\par\end{centering}
\centering{}\caption{The same as Figure \ref{cum_boots} but for
Mars' and its last impactor analogs.}
\label{fig:The-cumulative-mars}
\end{figure*}

\subsection{The composition of Mars' moons}

Mars' moons
were thought to be two captured asteroids, as indicated by their morphology and composition \citep{Mu91, Bu78}.
However, recent observations \citep[e.g.,][]{Gi11, Wi14} showed that their composition and density is hardy reconcilable with
this scenario and could be better understood if Phobos and Deimos were the
result of a giant impact with a body of mass a hundred times smaller than the mass of their
target \citep{Cra11, Cit15}.

We evaluated the cumulative distribution of the $\Delta^{17}O$ between 
Mars and its last impactor from both the raw simulations (see Figure
\ref{fig:mars_cum}) and from the bootstrapped sample (see Figure
\ref{fig:The-cumulative-mars}).

From the comparison between Figures \ref{fig:mars_cum} and \ref{fig:The-cumulative-mars} and Figures \ref{fig:The-cumulative-distribution} and \ref{cum_boots} 
{(and between Figures \ref{fig:cum_mars_mars_2au} and \ref{cum_mars_boots_mars2au} and Figures \ref{fig:cum_2au} and \ref{fig:cum_boots_mars2au}) } it is apparent that, if Mars' moons are the result of a giant impact, the differences between Mars analogs
and the relative last impactors are smaller than those found between the
planets formed in the system, but larger than the differences found for the Earth-Theia analogs.
In conclusion, we do not expect any extreme composition similarity between Phobos or Deimos and Mars,
however we predict a smaller difference respect to what would be expected in case of captured asteroids.

\begin{figure}
\begin{centering}
\includegraphics[width=0.5\textwidth, clip=true]{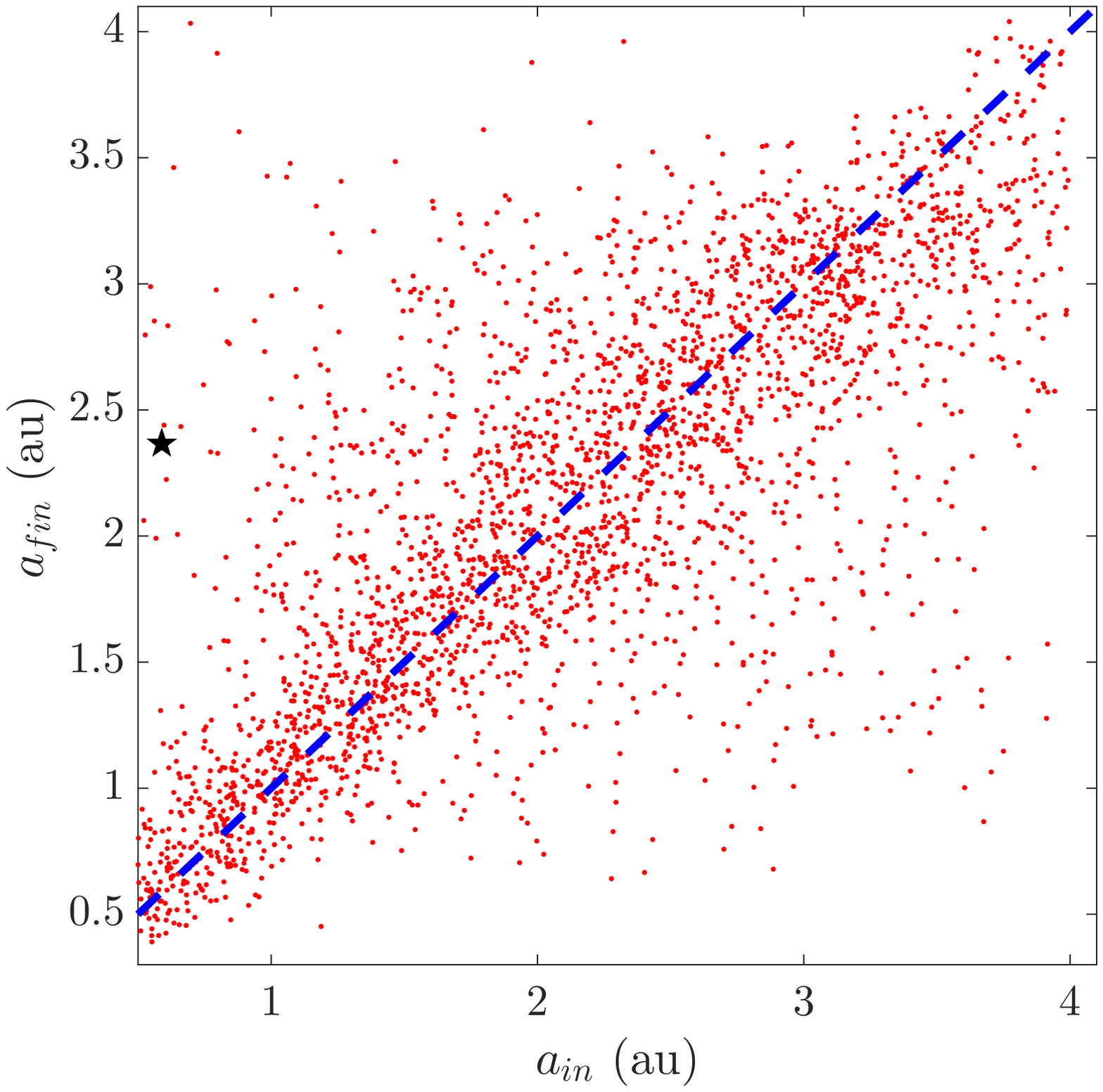}
\caption{The planetesimals that survive without impacting any planetary embryo until the end of each simulation with proper analogs. Their final position is plotted
against their initial position. Particles that did not move from their original position lay on the blue dashed line.  The black star shows the position of Vesta in the diagram, if its initial semi-major axis is calculated relying on a Solar System-based calibration.}
\label{fig:vesta}
\par\end{centering}
\end{figure}

\begin{figure}
\begin{centering}
\includegraphics[width=0.5\textwidth, clip=true]{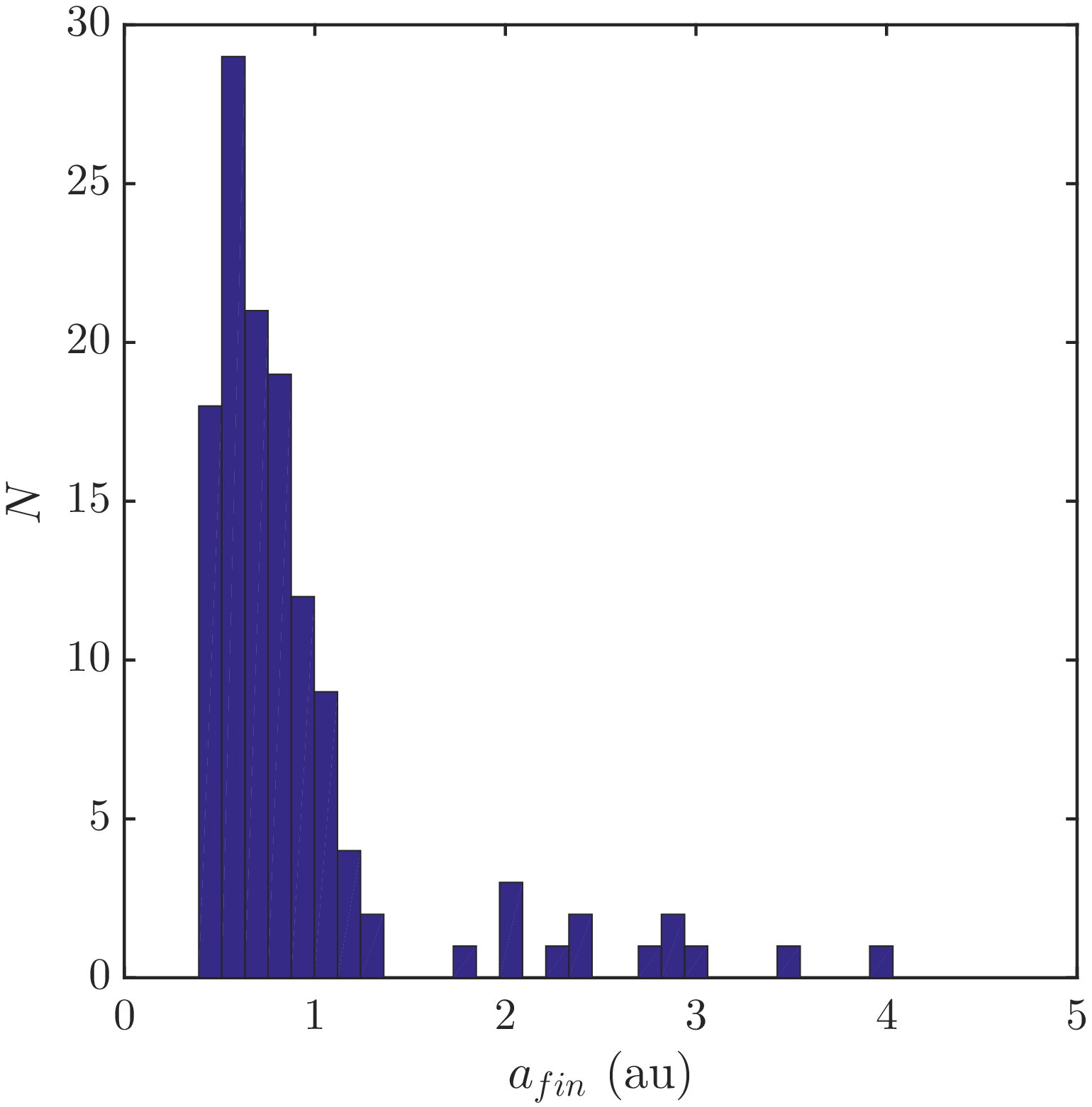}
\caption{Histogram showing the distribution of the final semi-major axis of the asteroids initially between $0.46$ and $0.72$ au, i.e. Vesta's initial radial range allowed by the error. Among all the asteroids, 8.7\% are scattered beyond $2$~au, i.e. in the asteroid belt, by the end of the simulation.}
\label{fig:vesta_hist}
\par\end{centering}
\end{figure}

\subsection{Solar system asteroids and the composition of Vesta}

Vesta is a large differentiated rocky asteroid
that formed during the first few million years of the Solar System, after the formation of the 
first solid planetesimals \citep{Ma12}. Its $\Delta^{17}O$
\citep[$-250\pm80$, compared to $0$ppm and $320$ppm for the Earth and Mars;][]{CM96, Fr99} suggest this asteroid did not form in the current position.
Since its isotopic composition does not match the composition gradient in the Solar
System, Vesta could have formed closer to, or farther from the Sun
to be then scattered on its current orbit {\citep[see][]{Bo06}}. 

To check this hypothesis we compared the initial and the final semi-major axis of the survived
planetesimals in the simulations that contain proper analogs (see Figure
\ref{fig:vesta}). The simulated planetesimals have a mass two orders
of magnitude larger than Vesta, however, they are the smallest particles
followed in our simulation and so they serve as the best reference for the asteroid distribution. 

Adopting the current position and compositions of the Earth and Mars to calibrate the Solar System $\Delta^{17}O$ gradient (see Eq. \ref{eqcal}) we obtain that Vesta had an initial semi-major axis $a_i=0.59\pm0.13$au (see black star in Figure \ref{fig:vesta}). If we consider all the asteroids, only a small fraction of them  (3\%) are scattered as much as expected for Vesta.
However, if we consider only the asteroids with an initial semi-major axis in the radial bin allowed for Vesta, i.e. between $0.46$au and $0.72$au, we find that
8.7\% of  these objects are significantly scattered in the asteroid belt, acquiring a semi-major axis larger than $2$~au by the end of the simulation (see Figure \ref{fig:vesta_hist}).
The scattering process in the early stages of the Solar System could then be one of the possible explanations for the existence of outliers like Vesta, beside predicting many other similar objects. 

\section{Summary and Conclusions}\label{sec:concl}
In this paper we used a large (140) sample of simulations of rocky planets formation 
to compare the composition of the Earth and Theia and the composition of Mars to
that of its last impactor. We also considered the case of Vesta, an asteroid whose position and composition
do not match. 
Here we summarise the main results of this work.
\begin{itemize}
\item {Considering  all the systems,  the raw simulation data yield a probability to get a moon 
with similar oxygen isotope composition to the Earth between 4\% and 14.5\% (0\%-4.6\% for the $6$ppm limit), depending on the percentage of 
mixing allowed (0 to 40\%) and $1\sigma$ Poissonian error. }
\item The analogs of the Earth and of  Theia are sistematically more similar than the planets.
\item To evaluate the effect of granularity affecting our simulations, we applied a bootstrapping technique with replacement
to all the simulations and we obtained new probabilities from the larger data sample obtained in this way. 
Including the possibility of mixing and considering the $1\sigma$ error, we find the probability of compatibility to be 4.9\%-12.2\% (1.9\%-5.2\% for $6$ppm),
when taking into account only systems with analogs formed by at least 5 particles and $3000$ bootstrapped samples. When considering a threshold of 50 particles composing each analog and bootstrapping 5000 times 
the probability is between 5.6\% and 18.2\% (2\%-6.7\% for $6$ppm). 
This fraction is somewhat lower than what found in Paper I for the reasons listed above, but generally confirms our previous results, showing that the 
composition similarity could arise naturally from the primordial composition similarity between the proto-Earth and Theia, hence potentially 
alleviating the main challenge to the giant-impact scenario. Moreover, even if the probability we find is comparable to the likelihood 
for a high mass-ratio impact ($6.7$\%) such as suggested by \cite{Ca12}, it does not require fine-tuned condition and extremely high-spin for the proto-Earth \citep{Cu12}.   
\item Mars' moons are expected to show a significantly different composition compared with their Mars-analog host. 
\item Asteroids, identified as the planetesimals that survive till the end of the simulation, can be scattered significantly far away their initial
position.  We found that 8.7\% of our leftover planetesimals with initial semi-major axis compatible, within the error, with the one evaluated for Vesta are farther from the Vesta linear expectation of non-migrating asteroids.  This process could explain the mismatch between Vesta distance from the Sun and its composition, as also predicted by \cite{Bo06}.
\end{itemize}

We conclude that, as already found by \cite{2015Icar..252..161K,2015Icar..258...14K},  the impact between enough similar planets is
rare; however this is the most probable and less fine-tuned mechanism  able to explain the formation of the Moon, its properties and composition. 
The giant-impact scenario, with its $\sim10\%$ probability to lead to a Earth-Moon similar composition is currently the most consistent model available.
If Mars' moons are the result of a giant impact it is highly improbable to
get to similar composition to their planet. Asteroids are not expected to closely follow the initial composition gradient
in the proto-planetary disk, since they could have been scattered from their initial position.
More detailed simulations are needed to further explore these issues and many other open questions related to our Solar System.

\section*{Acknowledgements}
We are grateful to Sean Raymond for his  helpful comments and discussions regarding this work.
We also  thank the referee, Nathan Kaib, for comments and
suggestions that improved the quality of this paper.
HBP acknowledge support from BSF grant number 2012384, Marie Curie FP7
career integration grant ``GRAND'', the Minerva center for life under extreme planetary
conditions and the ISF I-CORE grant 1829/12.

\bibliographystyle{mnras}
\bibliography{moon}
%%%%%%%%%%%%%%%%%%%%%%%
%%%%%%%%%%%%%%%%%%%%%%%

\appendix
\section{Mars' semi-major axis}\label{app:SMA}
\begin{center}
\begin{figure*}
\begin{centering}
\includegraphics[width=0.45\textwidth, clip=true]{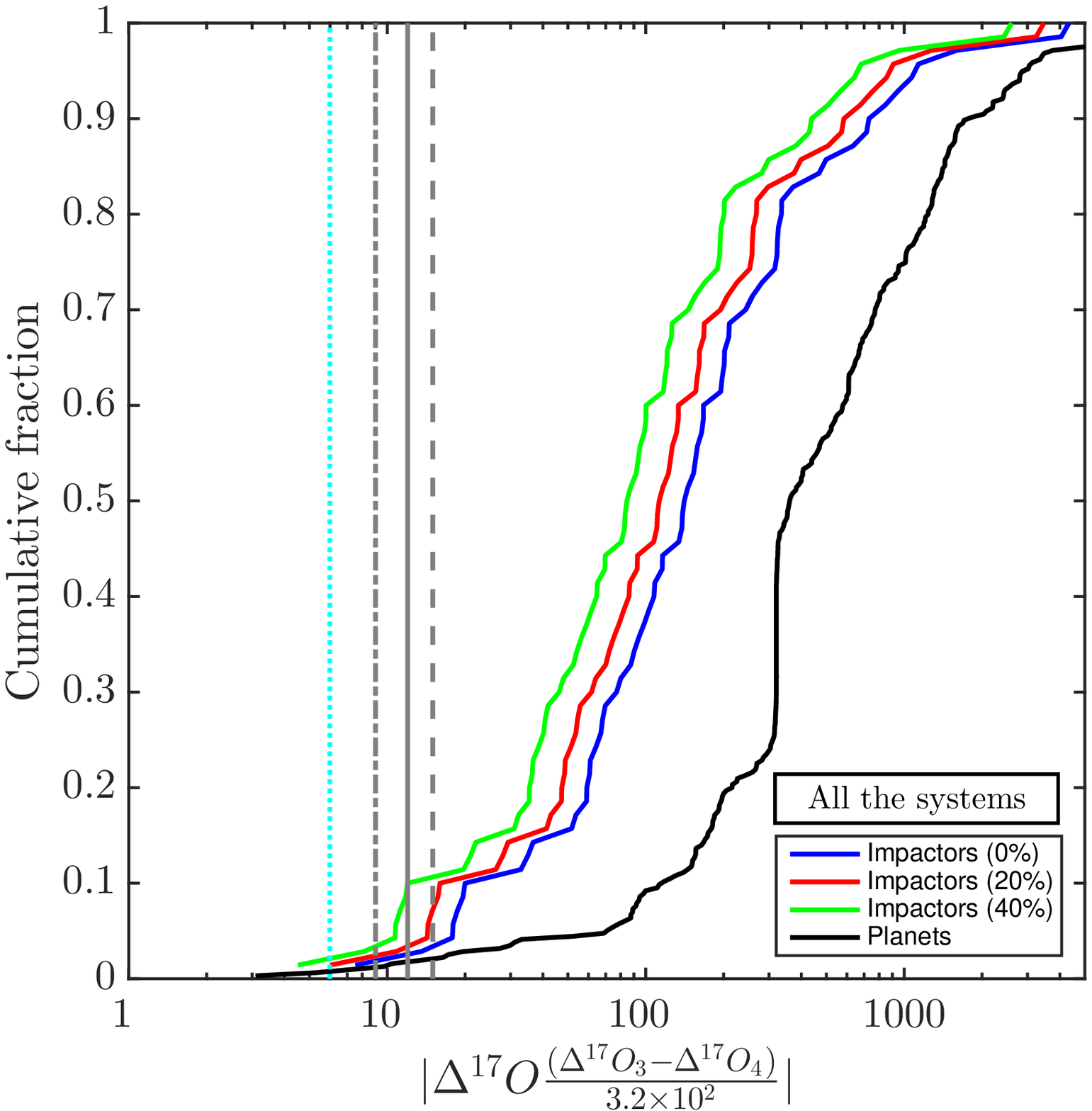}
\includegraphics[width=0.45\textwidth, clip=true]{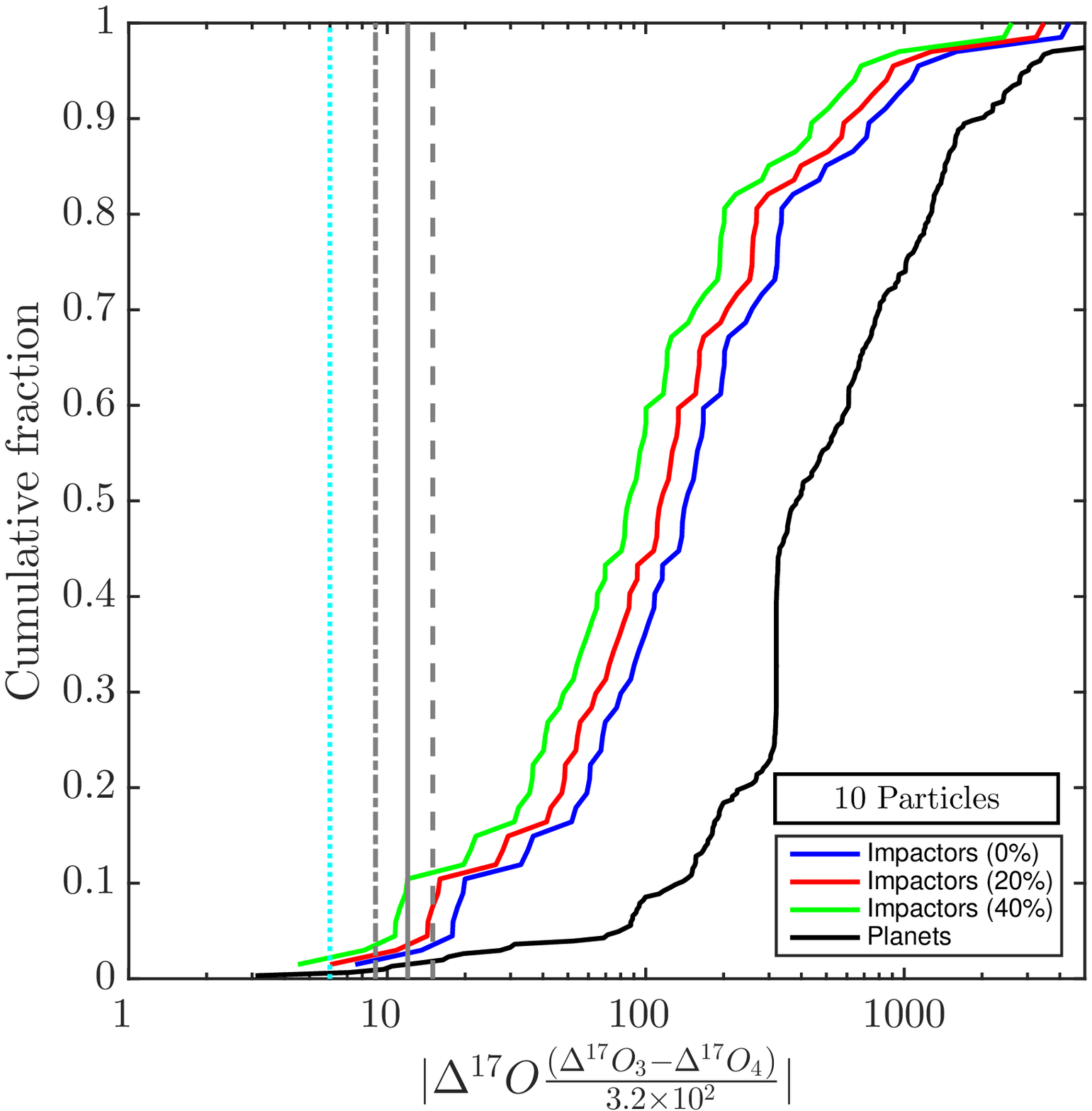}
\par\end{centering}
\centering{}\includegraphics[width=0.45\textwidth, clip=true]{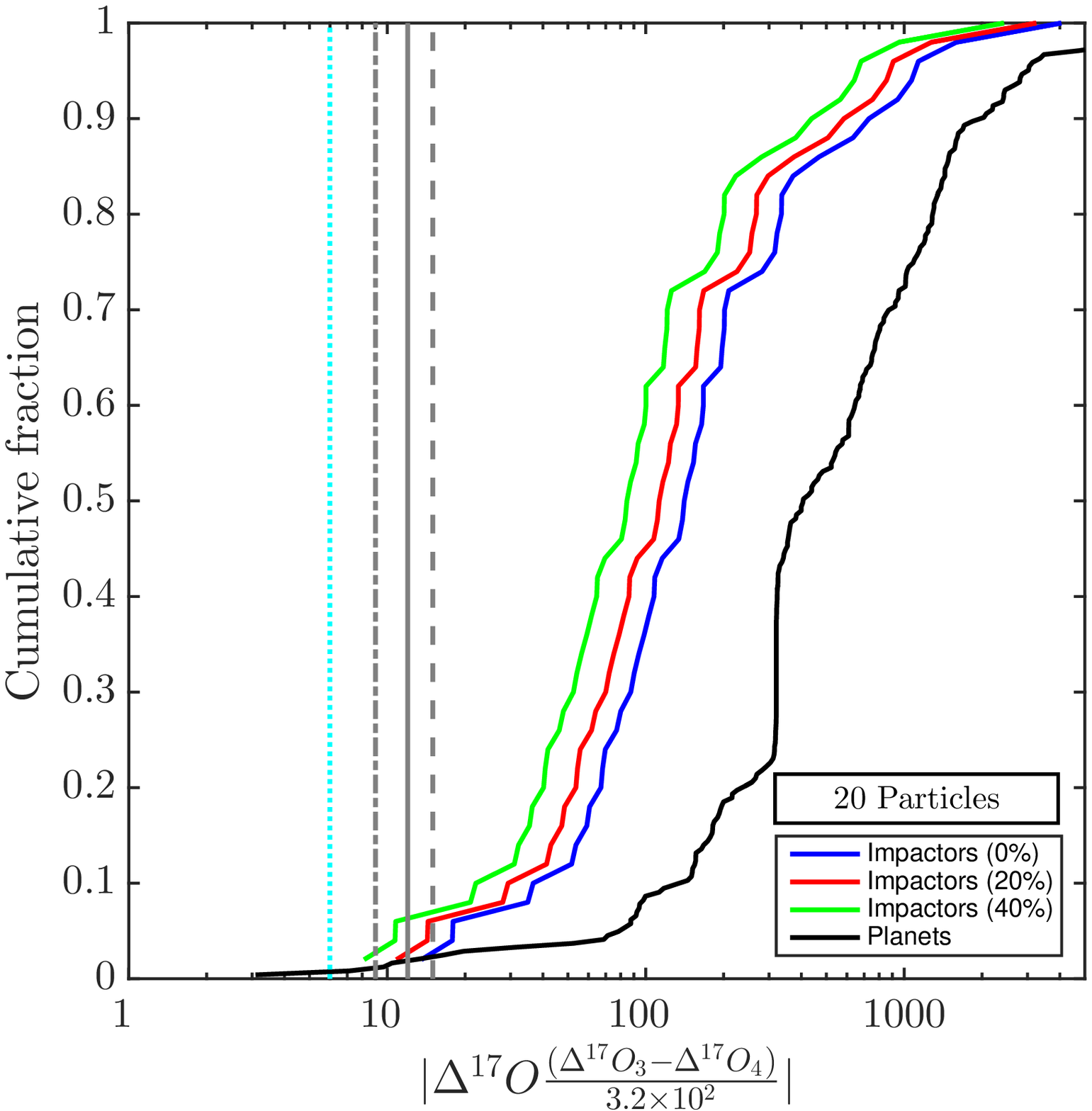}
\includegraphics[width=0.45\textwidth, clip=true]{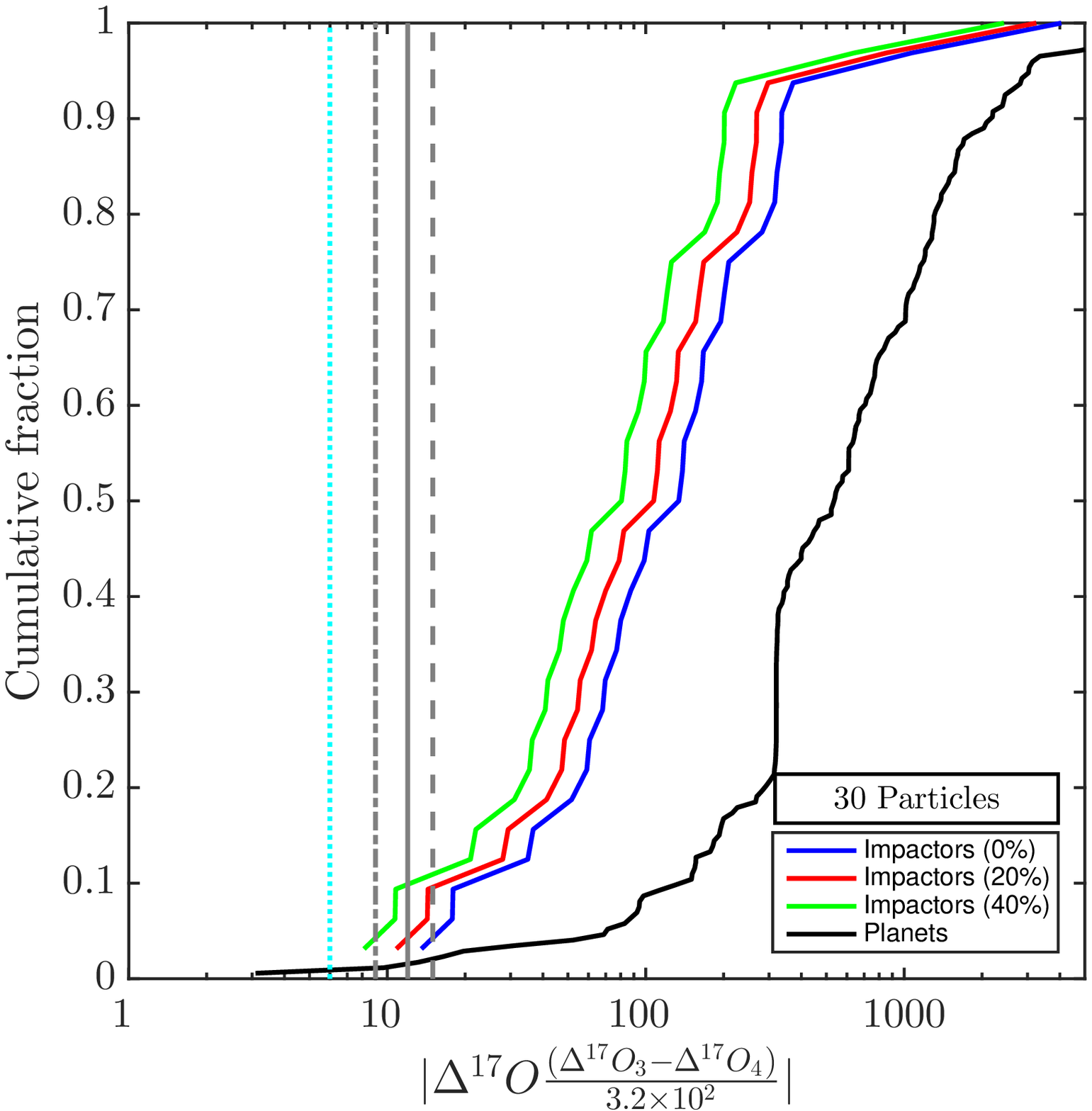}
\includegraphics[width=0.45\textwidth, clip=true]{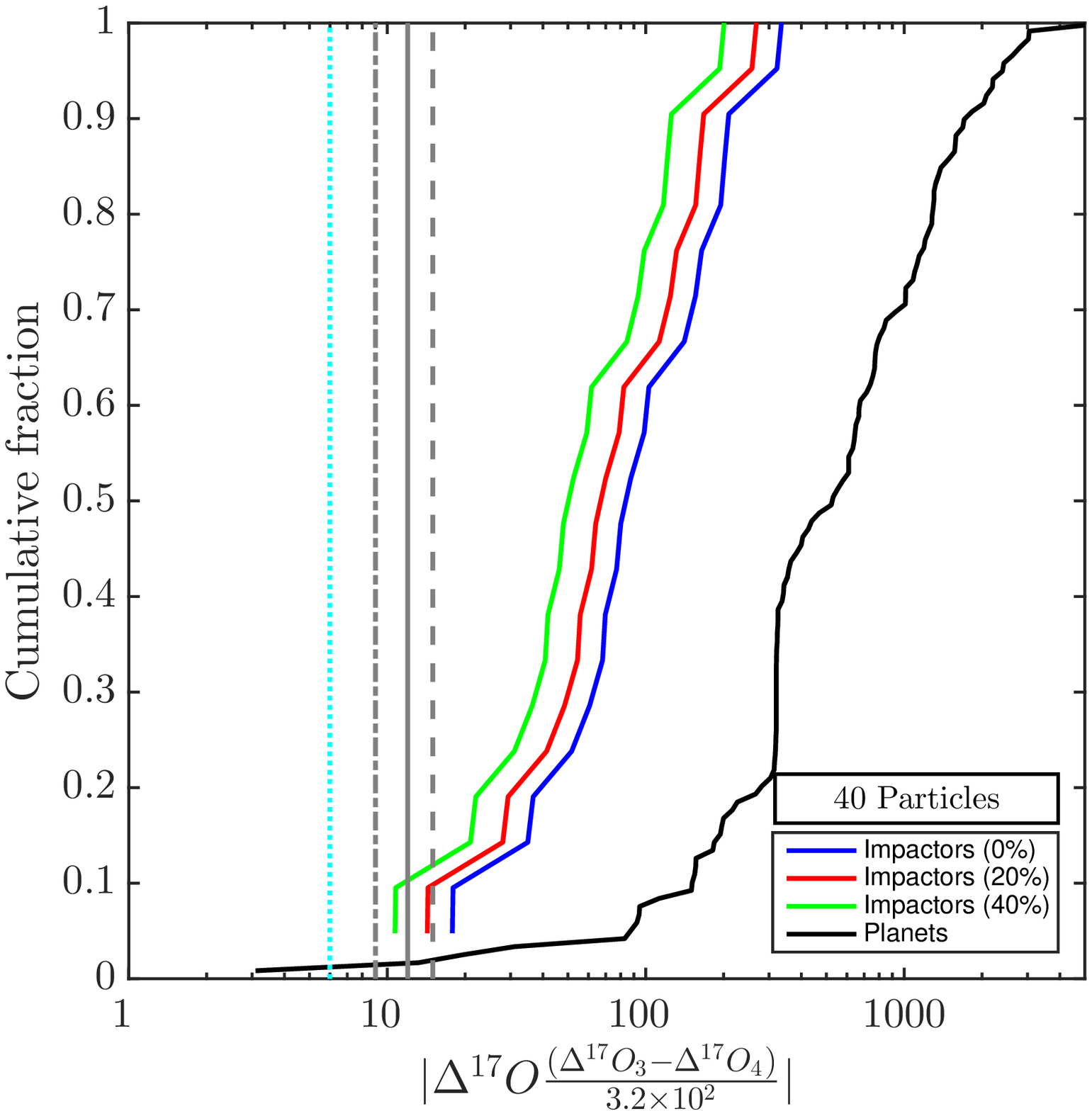}
\includegraphics[width=0.45\textwidth, clip=true]{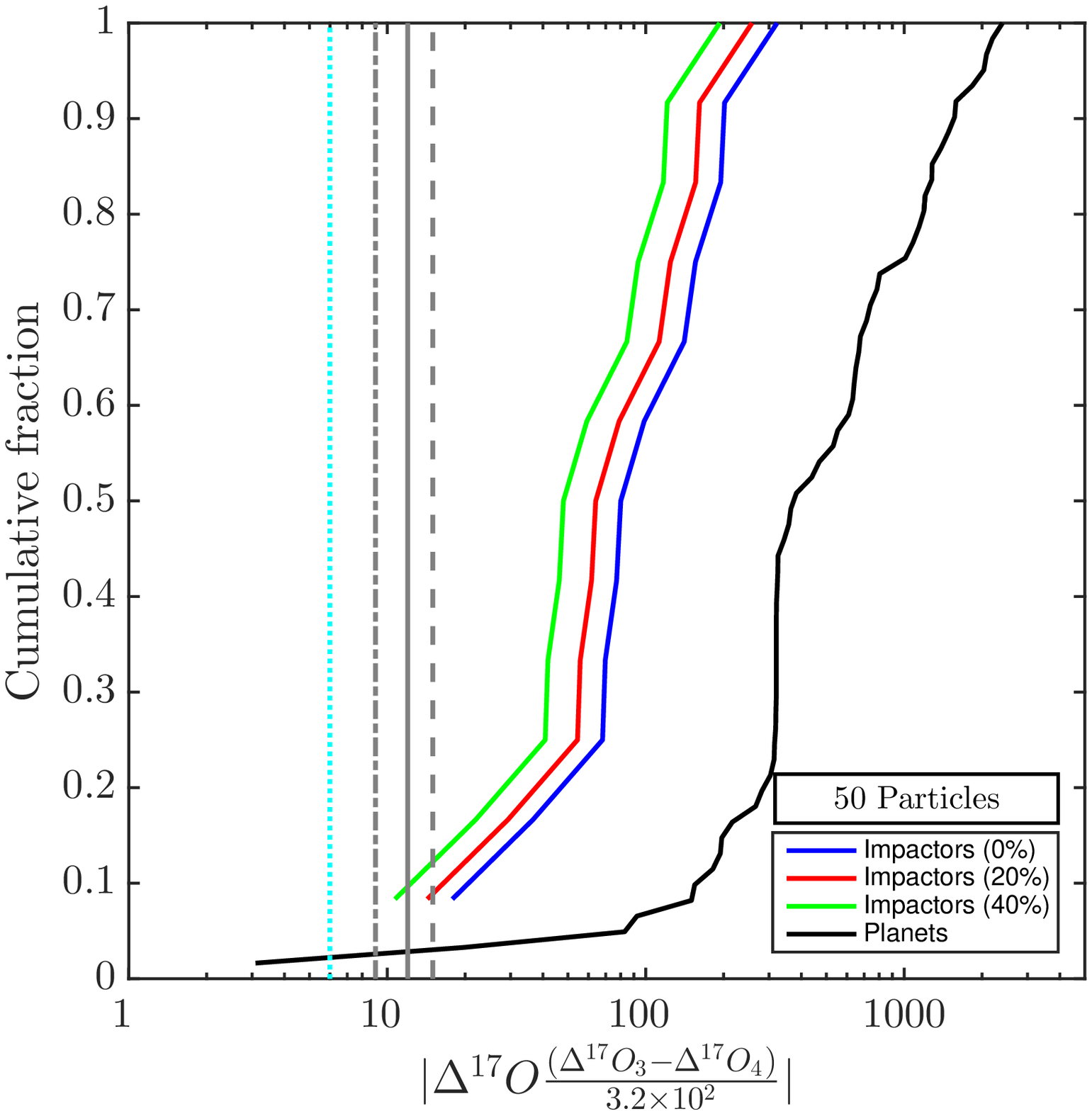}
\caption{The same as Figure \ref{fig:The-cumulative-distribution} obtained using an upper limit of $2$~au for the semi-major axis of Mars's analogs.}
\label{fig:cum_2au}
\end{figure*}
\end{center}

\begin{center}
\begin{figure*}
\begin{centering}
\includegraphics[width=0.5\textwidth]{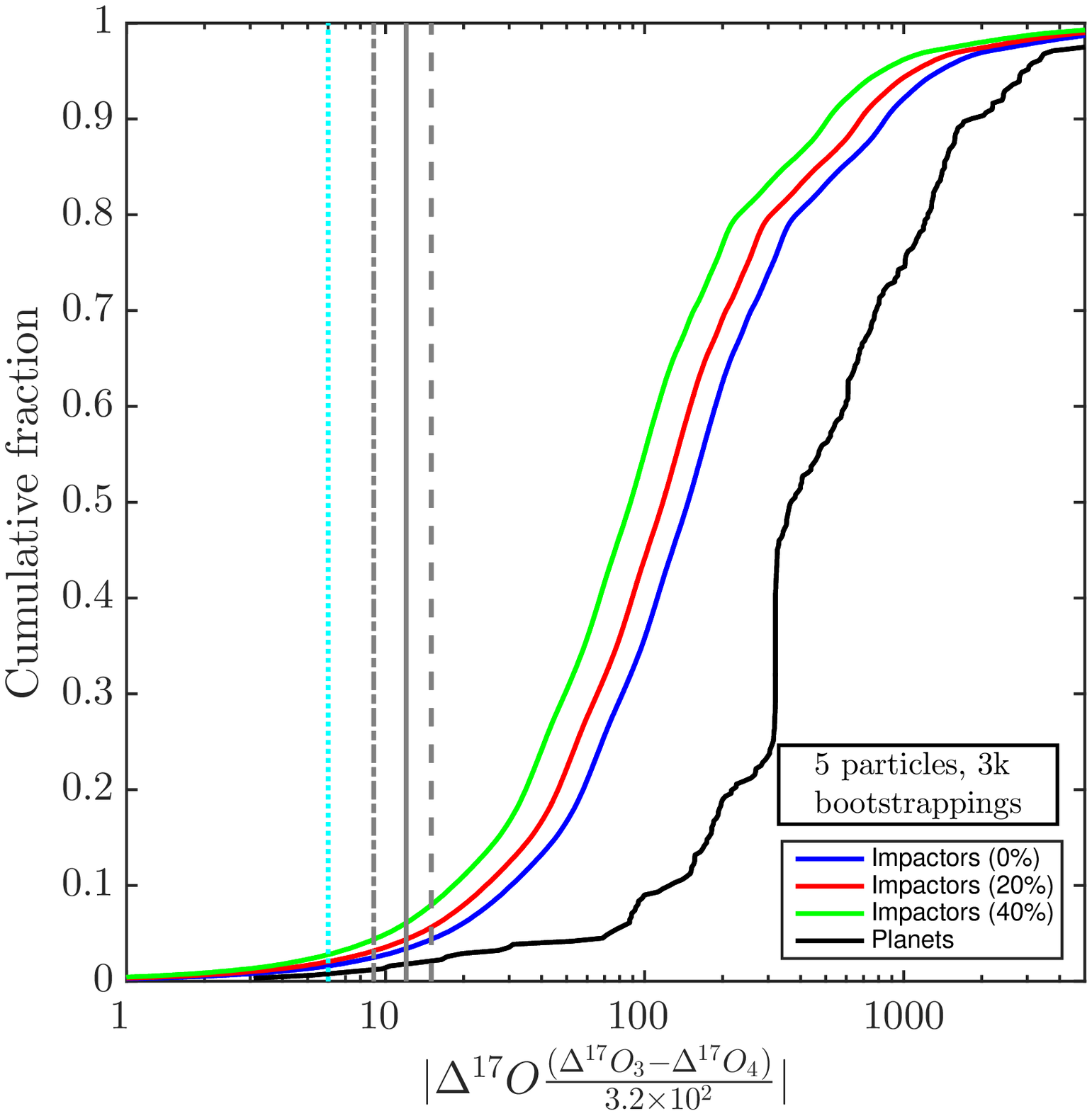}\includegraphics[width=0.5\textwidth]{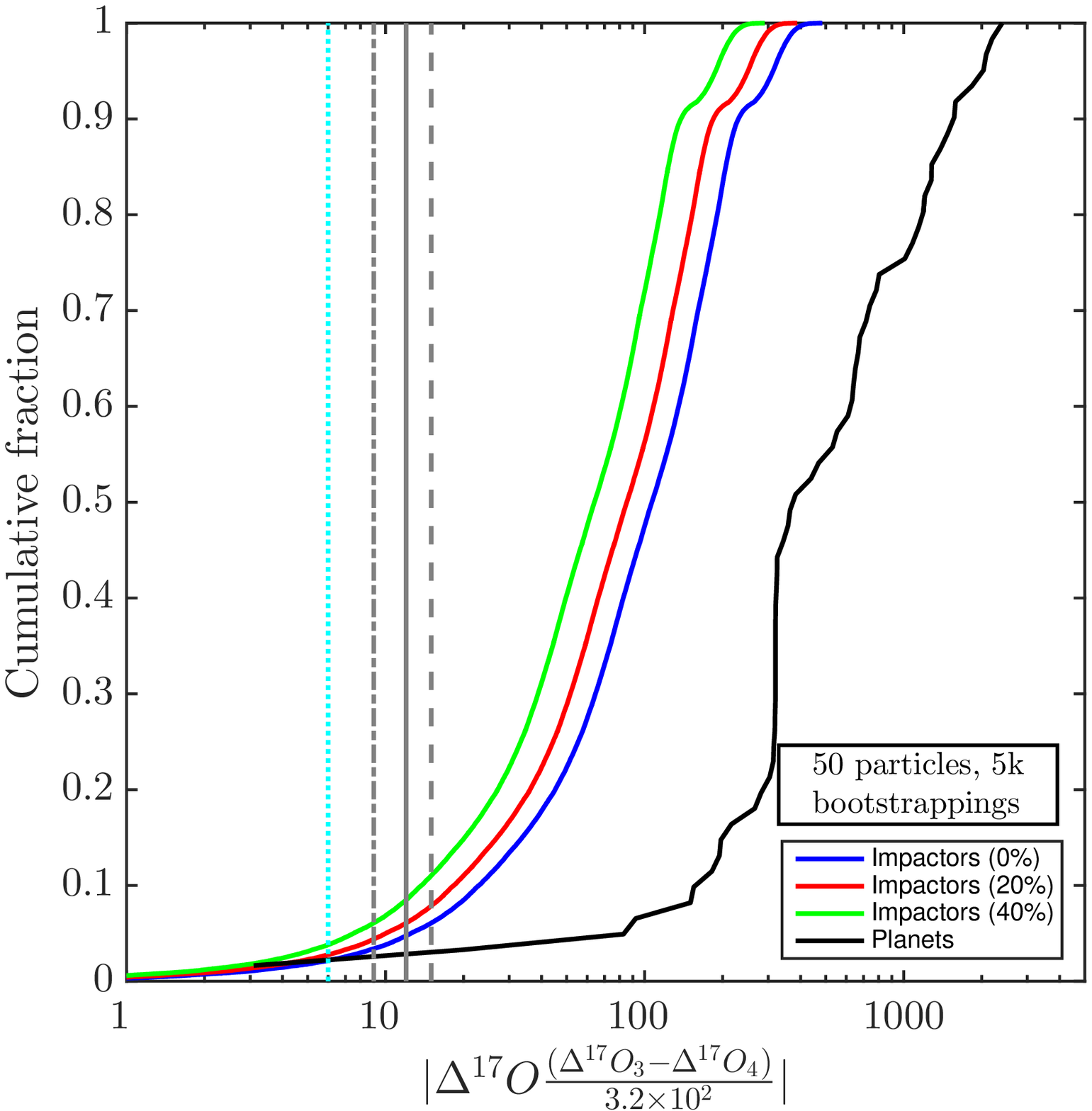}
\par\end{centering}
\caption{The same as Figure \ref{cum_boots} obtained using an upper limit of $2$~au for the semi-major axis of Mars's analogs.}
\label{fig:cum_boots_mars2au}
\end{figure*}
\end{center}

\begin{center}
\begin{figure*}
\begin{centering}
\includegraphics[width=0.45\textwidth, clip=true]{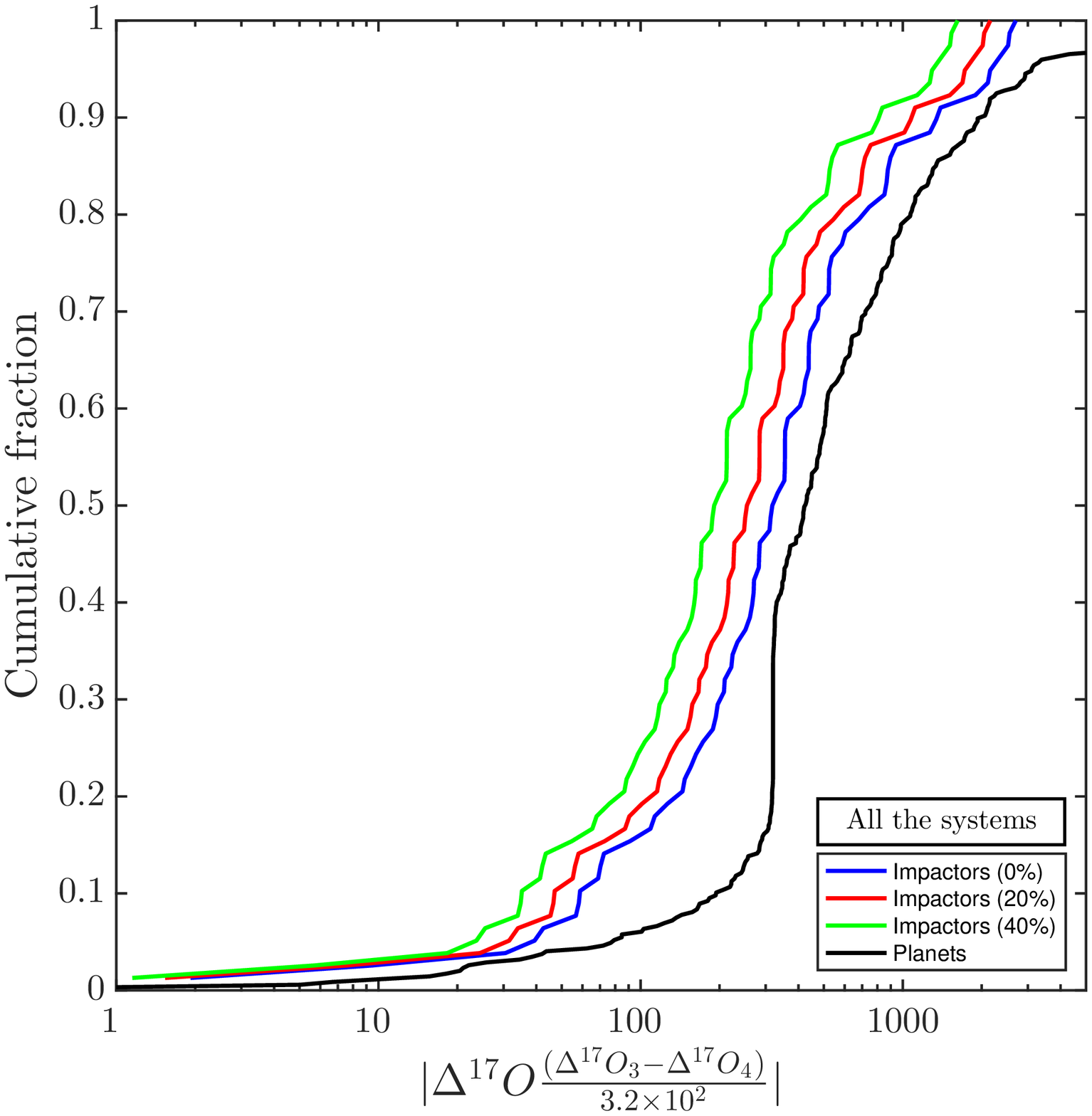}
\includegraphics[width=0.45\textwidth, clip=true]{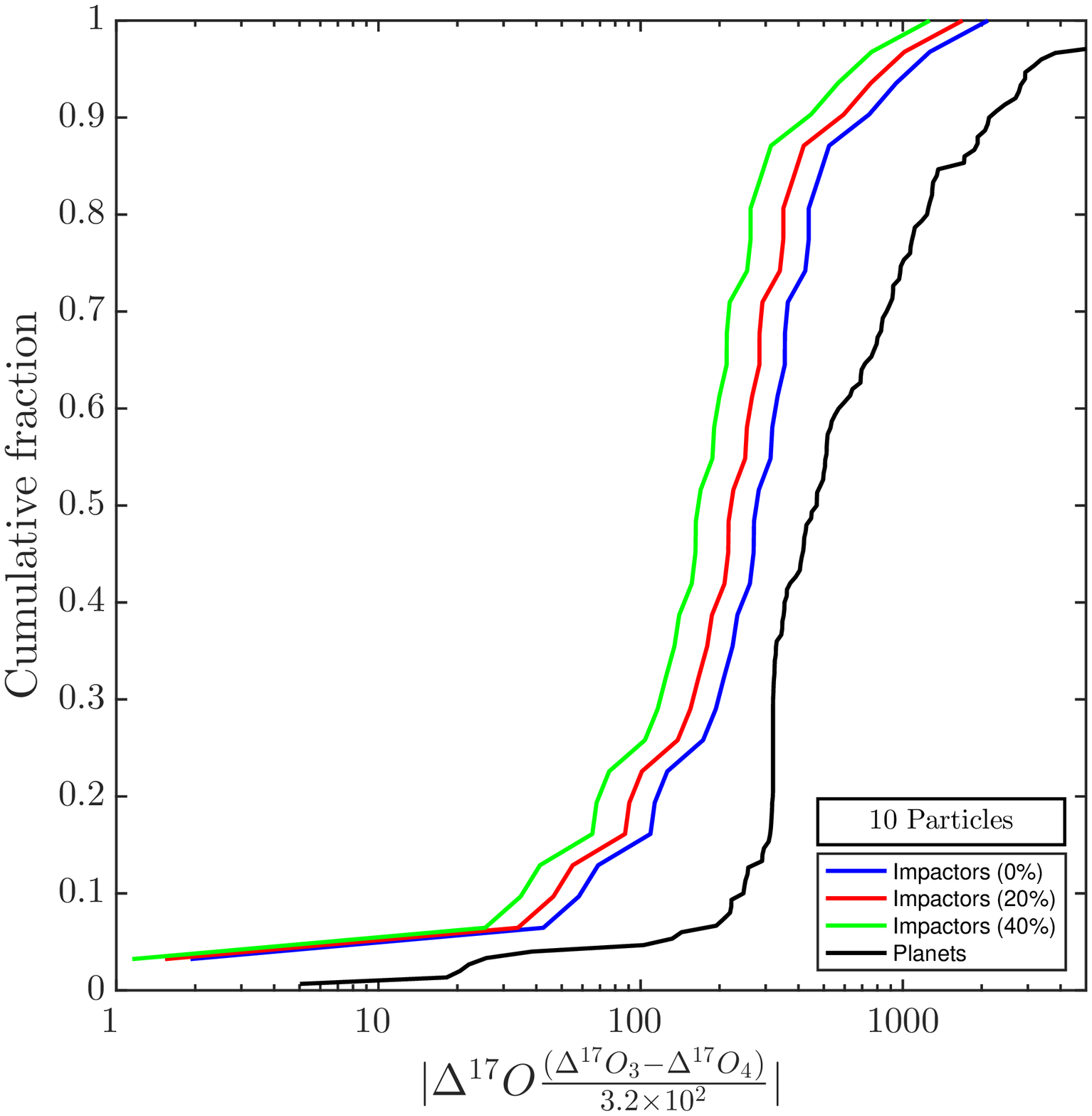}
\par\end{centering}
\centering{}\includegraphics[width=0.45\textwidth, clip=true]{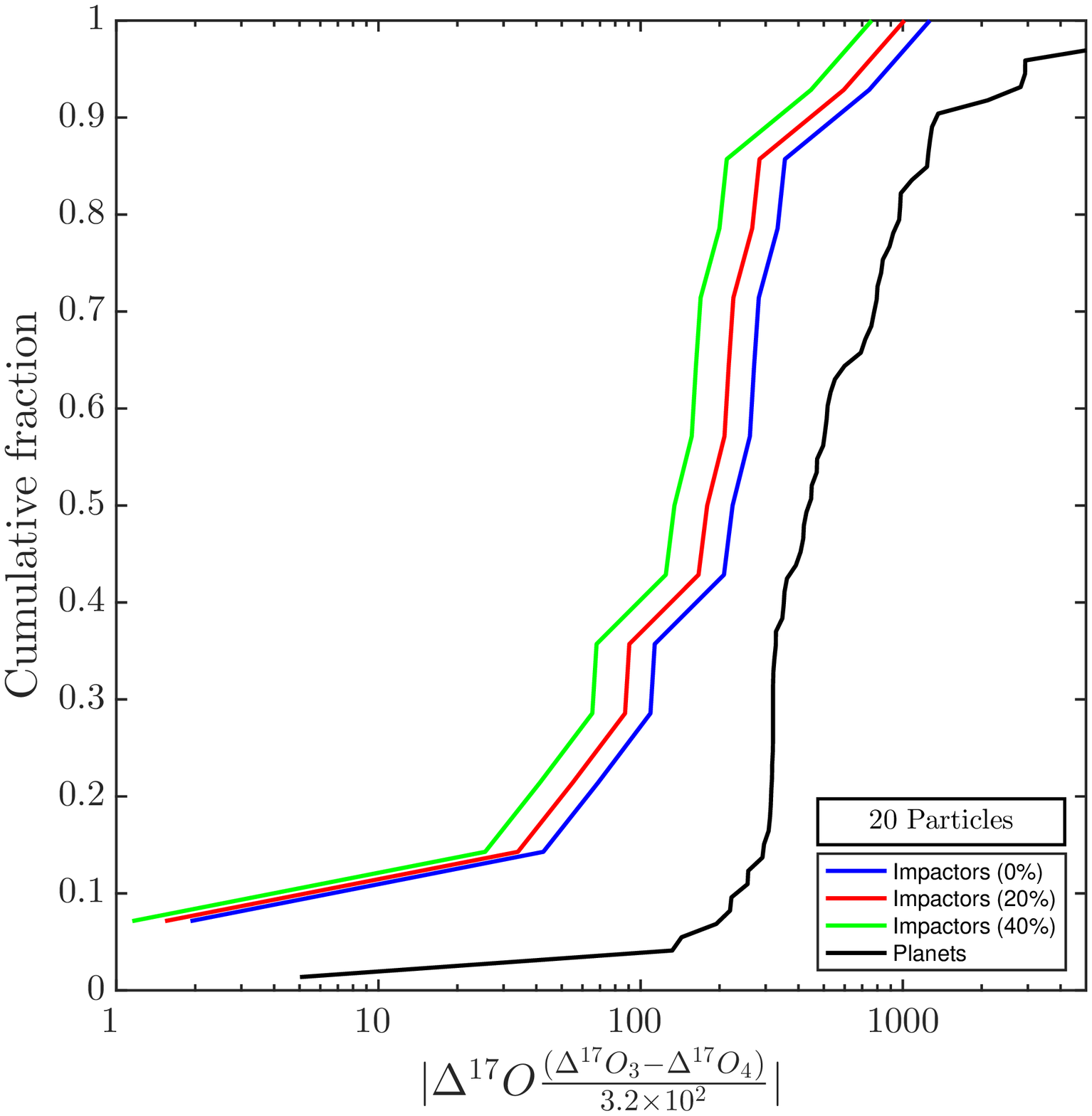}
\includegraphics[width=0.45\textwidth, clip=true]{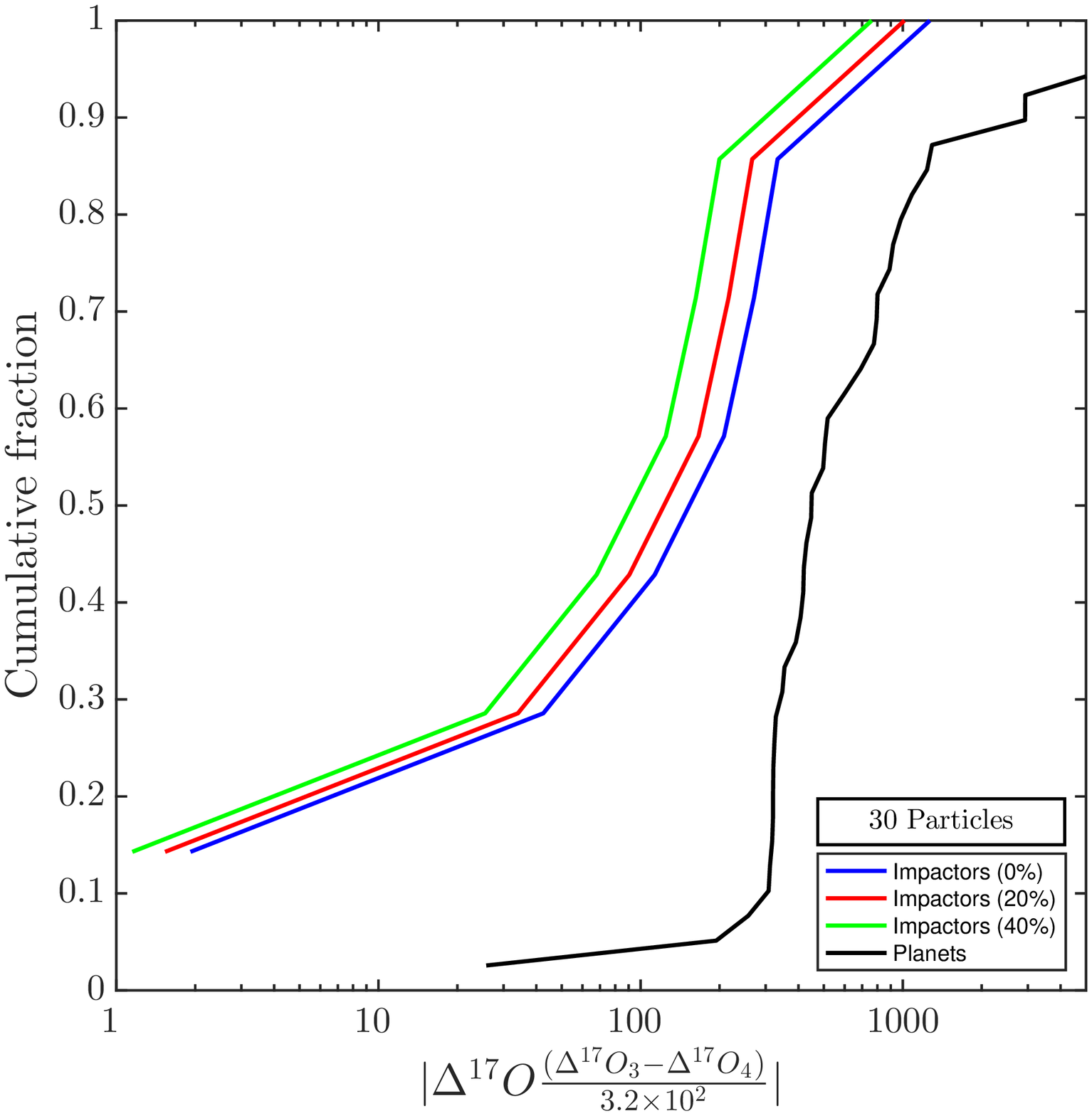}
\includegraphics[width=0.45\textwidth, clip=true]{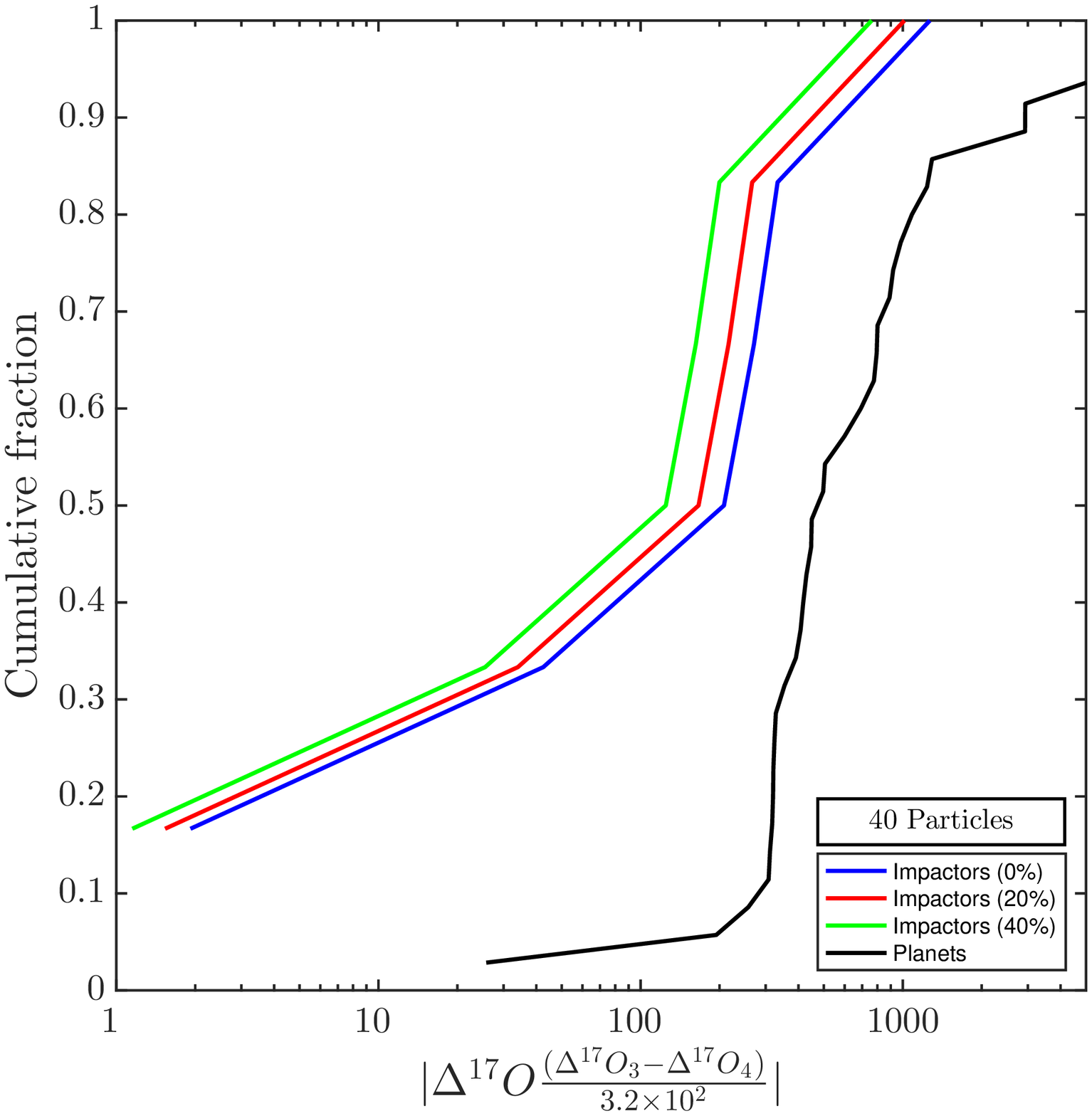}
\includegraphics[width=0.45\textwidth, clip=true]{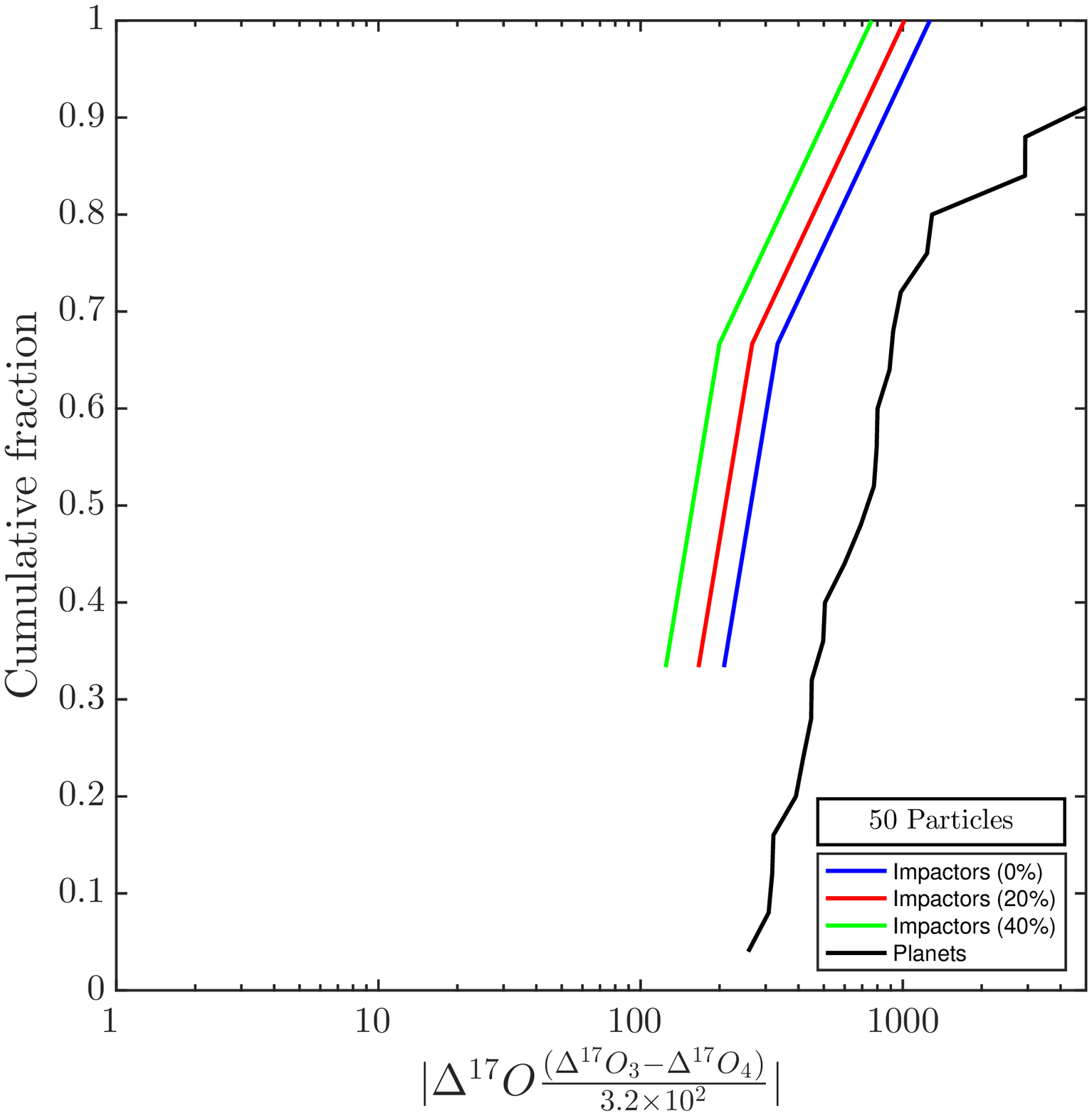}
\caption{The same as Figure \ref{fig:mars_cum} obtained using an upper limit of $2$~au for the semi-major axis of Mars's analogs.}
\label{fig:cum_mars_mars_2au}
\end{figure*}
\end{center}

\begin{center}
\begin{figure*}
\begin{centering}
\includegraphics[width=0.5\textwidth]{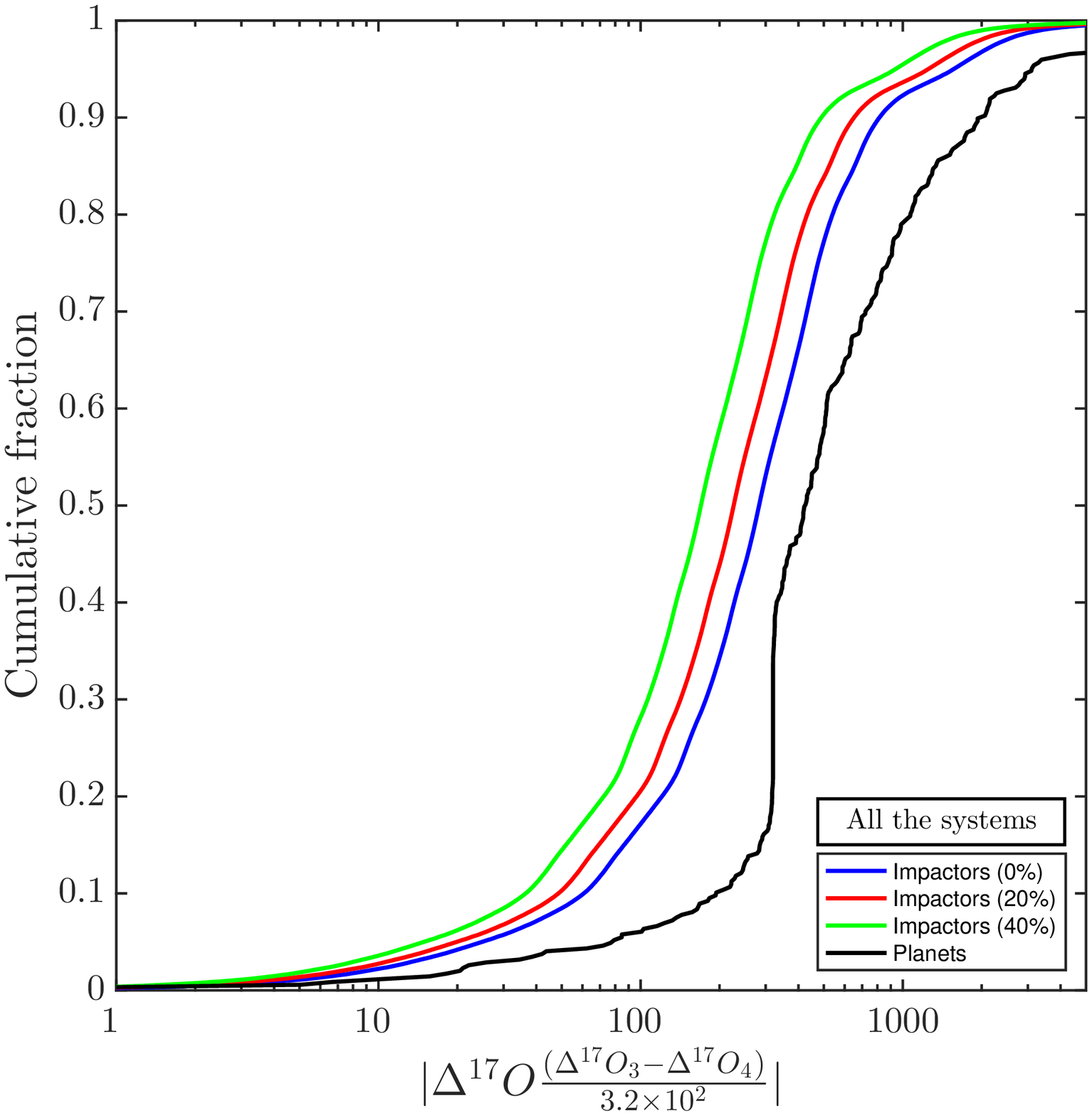}\includegraphics[width=0.5\textwidth]{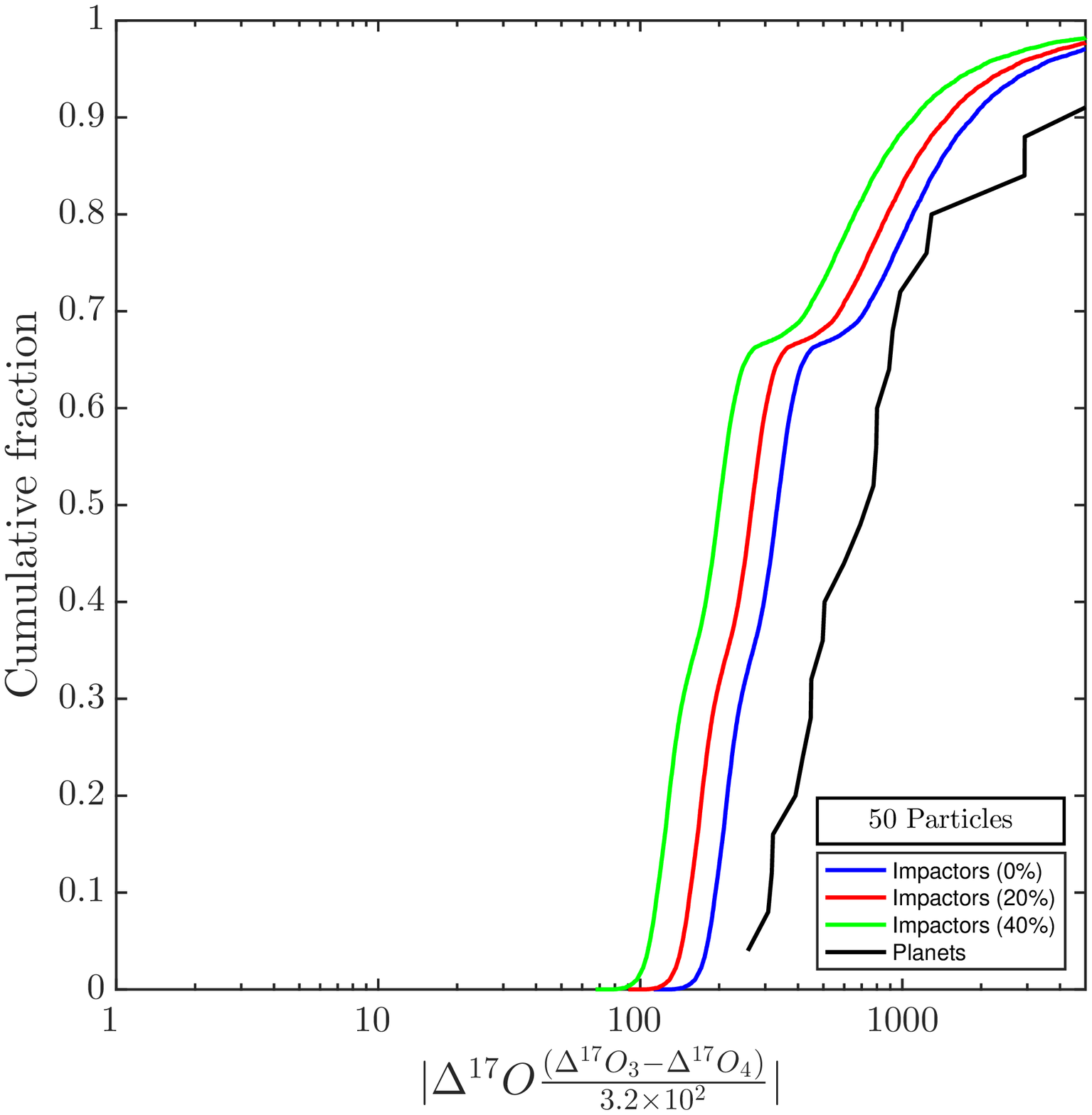}
\par\end{centering}
\caption{The same as in Figure \ref{fig:The-cumulative-mars} obtained using an upper limit of 2au for the semi-major axis of Mars's analogs.}
\label{cum_mars_boots_mars2au}
\end{figure*}
\end{center}

In order to check whether our choices affect the composition gradient of the proto-planetary disk,
we repeated our analysis setting an upper limit of $2$~au on the semi-major axis of Mars' analogs. 
As shown in Tables \ref{tab1} and \ref{tab2} there is no clear trend caused by this constraint
(see also Figures \ref{fig:cum_2au} and \ref{fig:cum_boots_mars2au}). \\
Figure \ref{fig:mass_ratio} does not change significantly and therefore, we are not showing its analog here.
The probability to have an
impact between two bodies with $\gamma>0.4$ is slightly smaller (6.5\%).\\
Repeating the analysis for Mars' moons we obtained comparable results to the ones obtained without limiting Mars' semi-major axis
(see Figures \ref{fig:cum_mars_mars_2au} and \ref{cum_mars_boots_mars2au}).\\
Vesta has the same probability to be a scattered asteroid; Figures \ref{fig:vesta} and \ref{fig:vesta_hist} are
unchanged.

\section{Step-function}\label{app:SF}
%%%%%%%%%%%
\begin{figure}
\begin{centering}
\includegraphics[width=0.5\textwidth, clip=true]{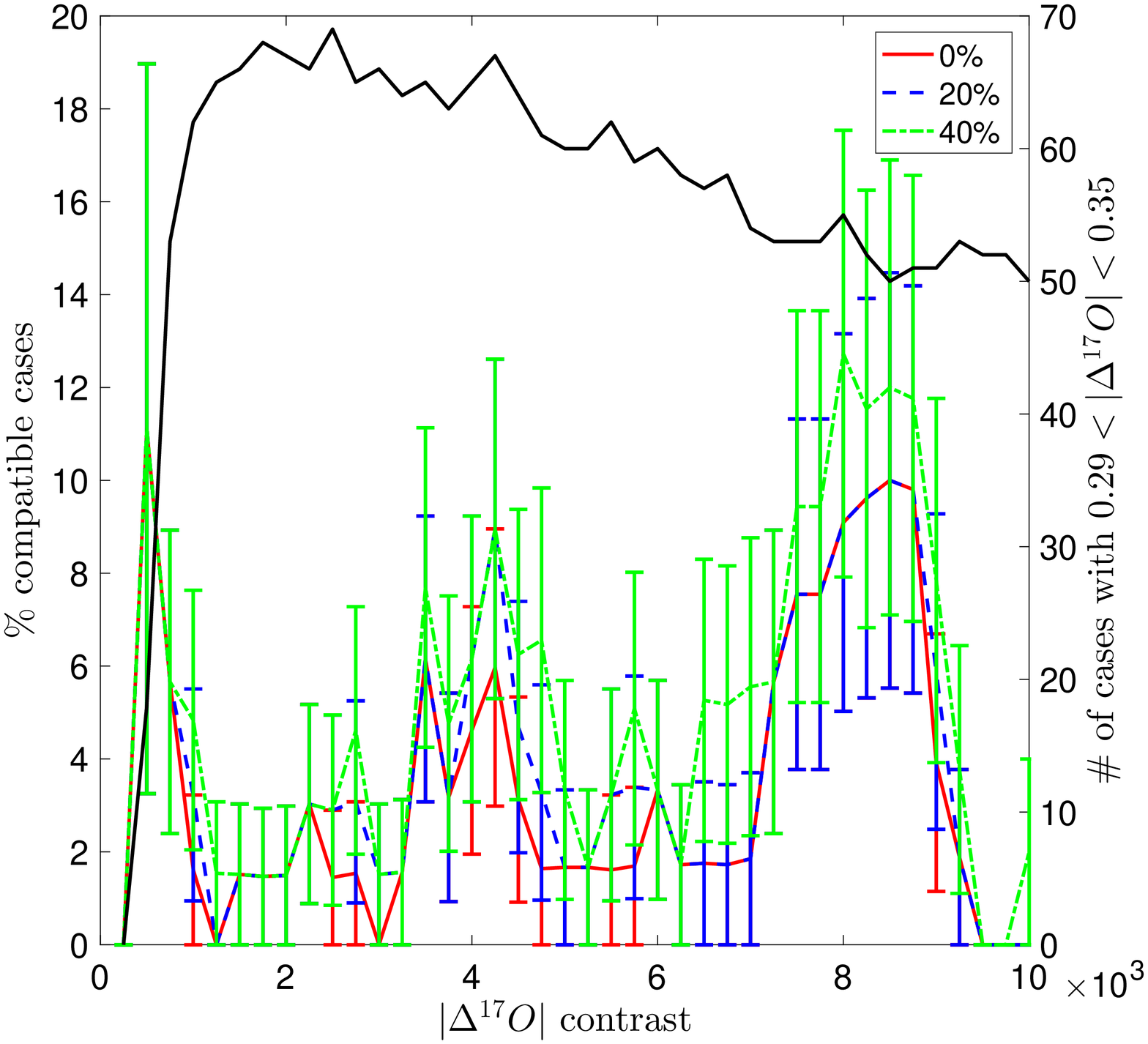}
\caption{The $\Delta^{17}O$  difference between the Earth's and Theia's analogs, and the relative $1\sigma$ poissonian error, are shown as a function of the contrast between the inner and outer proto-planetary disk for mixing fractions of 0\%, 20\% and 40\% (red solid, blue dashed and green dot dashed lines).  The black solid line is the number of cases in which the composition difference between the Earth and Mars is found to be in the correct range. }
\label{fig:step_fun}
\par\end{centering}
\end{figure}
%%%%%%%%%%%
The composition gradient of the proto-planetary disk is unknown; the assumption of a linear gradient done in Section \ref{cal} is then only
one of the possible functions that can be used to model the $\Delta^{17}O$ distribution.
Many functions have been used in literature \citep[see][]{2015Icar..252..161K, 2016Sci...351..493Y}, however, according 
to the recent results presented by \cite{Dau17}, the composition of the proto-planetary disk was not random, 
and the Earth and Theia aggregated mostly from planetesimals with a similar composition.
Using lithophile, as well as moderately and highly siderophile elements to trace the accretion history of the Earth,
\cite{Dau17} found that the inner proto-planetary disk had a homogeneous composition, dominated by enstatite chondrites, while
the outer part contained a larger fraction of ordinary chondrites. 
Therefore, we tested the giant-impact scenario assuming a sharp contrast  between the composition of inner and outer
terrestrial planets modelled as a step function for the $\Delta^{17}O$ distribution.
As done by \cite{2015Icar..252..161K} we initially fixed a $\Delta^{17}O$ contrast and then evaluated the heliocentric distance
at which the change has to occur in order to have a $\Delta^{17}O$ between Mars and the Earth of
$320\pm30$ppm.
We used contrasts between 0 and $10^4$ppm \citep{2015Icar..252..161K, Bur04, Ozi07} and we evaluated the fraction of compatible
Earth-Theia couples in each case. We assumed an upper limit of $2$~au for Mars' analogs semi-major axis.
The  number of cases for which it is possible to approximately reproduce the composition
difference between Mars and the Earth (black solid line in Figure \ref{fig:step_fun}) peaks around a contrast of $2000$-$4000$ppm and produces 50-70 cases 
for larger contrasts. 
The fraction of positive cases varies significantly with the contrast and the fraction of mixing, ranging
between 0\% and 19\% when considering the 1$\sigma$ Poissonian error.
Therefore, using a step function for the composition gradient can produce a significant number of compatible 
impacts, comparable to what obtained when adopting a linear distribution.

\end{document}